\documentclass[manuscript]{aastex}

\usepackage{lscape}

\shorttitle{R Coronae Borealis stars}
\shortauthors{Hema et al.}

\begin{document}

\title{The Galactic R Coronae Borealis stars\,: the C$_2$ Swan bands, \\
   the carbon problem, and the $^{12}$C/$^{13}$C ratio}

\author{B. P. Hema$^{1}$, Gajendra Pandey$^{1}$, and David L. Lambert$^{2}$}
\affil{$^{1}$ Indian Institute of Astrophysics, Bangalore
    Karnataka 560034, India; hema@iiap.res.in, pandey@iiap.res.in}
\affil{$^{2}$ The W.J. McDonald Observatory, University of Texas at Austin,
Austin, TX 78712-1083, USA; dll@astro.as.utexas.edu}

\begin{abstract}

Observed spectra of R Coronae Borealis (RCB) and hydrogen-deficient carbon (HdC) stars are analyzed by synthesizing the C$_2$ Swan bands (1,0), (0,0), and (0,1)
 using our detailed line list and the Uppsala model atmospheres. The (0,1) and 
(0,0) C$_{2}$ bands are used to derive the $^{12}$C abundance, and the (1,0) 
$^{12}$C$^{13}$C band to determine the $^{12}$C/$^{13}$C ratios.
The carbon abundance derived from the C$_2$ Swan bands is about 
the same 
for the adopted models constructed with different carbon abundances over the 
range: 8.5\,(C/He\,=\,0.1\%), to 10.5\,(C/He\,=\,10\%). Carbon abundances 
derived from C\,{\sc i} lines are about a factor of 4 lower than the carbon 
abundance of the adopted model atmosphere over the same C/He interval, as 
reported by Asplund et al. (2000), who dubbed the mismatch between adopted and 
derived C abundance the `carbon problem'.
In principle, the carbon abundances obtained from C$_2$ Swan bands and that 
assumed for the model atmosphere can be equated for a particular choice of 
C/He that varies from star to star. Then, the carbon problem for C$_2$ bands 
is eliminated. However, such C/He ratios are in general less than those of the 
EHe stars, the seemingly natural relatives to the RCB and HdC stars. A more 
likely solution to the C$_2$ carbon problem may lie in a modification of the 
model atmosphere's temperature structure.
The derived carbon abundances and the $^{12}$C/$^{13}$C ratios are discussed in light of the double degenerate (DD) and the final flash (FF) scenarios.

\end{abstract}

\keywords{Stars - abundances, Isotopic ratio- carbon, evolution of stars}

\section{Introduction}
R Coronae Borealis (RCB) stars are a rare class of F- and G\,-\,type supergiants
with remarkable
photometric and spectroscopic peculiarities.
The photometric peculiarity is that a RCB may fade rapidly in visual brightness by
up to several magnitudes at unpredictable times
and slowly return back to maximum light after an interval of weeks, months or
even years.
Most RCB stars stay for a longer time at maximum light than at  minimum light.
This fading is generally attributed to the formation of dust in the line of sight.
Spectroscopic peculiarities are led by the  very weak or undetectable
hydrogen Balmer
lines in their
spectra. This indicates that they have a very H\,-\,poor atmosphere.
This hydrogen deficiency but not the propensity to undergo optical declines is
shared by other
rare classes of stars\,:\,extreme helium (EHe) stars at the hotter end and
hydrogen-deficient carbon (HdC) stars at the cooler end of the RCB temperature range.

Keys to understanding  origins of RCB stars and their putative relatives have
come from the determination and interpretation of the stars' surface chemical
compositions.
Two proposed scenarios remain in contention.
 In one dubbed the double degenerate (DD) scenario, a helium white dwarf merges with a
carbon-oxygen (C-O) white dwarf
\citep{webbink84,iben85}.
The close white dwarf binary results from mass exchange and mass loss of a binary
system as it evolves from a pair of main sequence stars. The final step to the
merger is driven
by  loss of angular momentum by
gravitational waves \citep{renzini79}. The envelope of the merged star is inflated to
supergiant dimensions for a brief period.
An alternative scenario dubbed the final flash (FF) scenario involves
a single post-AGB star experiencing a final helium
shell flash which causes the H-rich envelope to be ingested by the He shell.
The result is the star becomes a hydrogen-deficient supergiant for a brief period,
and is sometimes  referred in this condition as a born-again AGB star
\citep{renzini90}.

For the RCB stars,  determination of chemical compositions by
\citet{lambert94}  and \citet{asplund00}  suggested that the DD rather
than the FF scenario gave the superior accounting of the determined elemental
abundances. This conclusion has since been supported by the determination
from analysis of CO infrared bands of a high $^{18}$O (relative to $^{16}$O) in cool
RCBs and HdC stars
\citep{clayton05,clayton07, garc09, garc10}. Additional
evidence comes from high fluorine abundances in EHe \citep{pandey06} and
RCB stars \citep{pandey08}.

In the case of the RCB stars, there is an unease about the results for the elemental
abundance on account of `the carbon problem' identified and discussed by
\citet{asplund00}. Since the continuous opacity in the optical is predicted to
arise from the photoionization of neutral carbon from highly excited states, the
strength of an optical C\,{\sc i} line, also from a  highly excited state, is predicted
to be quasi-independent of atmospheric parameters such as effective temperature,
surface
gravity and metal abundance. Indeed, a C\,{\sc i} line has a nearly
constant strength across the RCB sample even as (for example) a Fe\,{\sc i} or
Fe\,{\sc ii} line may vary widely in strength from one star to the next. However,
the predicted strength of
a C\,{\sc i} line is much stronger than its observed strength: if one were to 
choose to resolve this discrepancy by adjusting the line's  $gf$-value, 
it must
be reduced by a factor of four or 0.6 dex on average. This discrepancy between
predicted and observed C\,{\sc i} line strengths is termed `the carbon problem'.
Adjustment of the $gf$-values of the C\,{\sc i} lines is not the only potential
or even the preferred way to address the carbon problem.

In this paper, we present and discuss spectra showing the C$_2$ Swan bands in
a sample of RCB and HdC stars. Our first goal is to compare predicted and observed
strengths of C$_2$ Swan
bands in RCB stars to see if they exhibit a carbon problem and if that
problem differs from that shown by the C\,{\sc i} lines.
Our second goal is to look for $^{12}$C$^{13}$C lines and determine the
$^{12}$C/$^{13}$C
ratio.
A high value of $^{12}$C/$^{13}$C ratio is expected for the DD scenario, but a low
ratio seems likely for the FF scenario.
High ratios or high lower limits on the isotopic ratio have been
set for
HdC stars\,:\,HD\,137613 \citep{fujita77}, HD\,182040
\citep{climenhaga60, fujita77}.
A limit of greater than 40 was set for R\,CrB \citep{cottrell82}.
But the RCB star V\,CrA is apparently an exception with a reported low value of
$^{12}$C/$^{13}$C ratio: \citet{rao08} estimated the ratio
at  4\,-\,10 for V\,CrA.

As expected, a low value of $^{12}$C/$^{13}$C ratio is shown by the final
flash object V4334\,Sgr (Sakurai's object), the ratio is 2 to 5
\citep{1997A&A...321L..17A, pavlenko04}. However, the other objects which are thought to be
final flash objects, like, FG\,Sge \citep{gonzalez98} and V605\,Aql
\citep{lundmark21, clayton97}, do not show the presence of
$^{12}$C$^{13}$C bands in their spectrum.

\section{Observations}

High resolution optical spectra of RCB/HdC stars at maximum light were
obtained from the W. J. McDonald Observatory and the Vainu Bappu Observatory.
The dates of observations,  the visual validated magnitudes
(AAVSO\footnote{http://www.aavso.org}) and the signal-to-noise
ratio per pixel of the spectra in the continuum near the 4737\,\AA\
 $^{12}$C$_2$ bandhead  are given
in Table\,1.
In addition to the RCB stars, a spectrum of $\gamma$\,Cyg was obtained at the
McDonald Observatory. This F5Ib star is of similar spectral type to the
warm RCBs such as R\,CrB.

The spectra from the McDonald Observatory were obtained with the 2.7-m
Harlan J. Smith Telescope and Tull coud\'{e} cross-dispersed echelle
spectrograph \citep{tull95} at a resolving power of
$\lambda$/d$\lambda$ = 60,000. The spectra from
the Vainu Bappu Observatory were obtained with the 2.34-m Vainu Bappu Telescope (VBT)
equipped with the fiber-fed cross-dispersed echelle spectrometer
\citep{rao05} and a 4K\,$\times$\,4K CCD are at a resolving power
of about 30,000.

\section{Spectrum synthesis}

Our analysis of the high-resolution spectra proceeds by fitting synthetic
spectra to the observed spectra in several bandpasses providing lines of the
C$_2$ Swan system.
For the synthesis of the C$_2$ Swan bands, we use model atmospheres and as
complete a line list as possible. In the following subsections we
introduce the line lists for the C$_2$ Swan bands and the atomic lines blended with
the C$_2$ bands and, finally, the procedure for computing the synthetic
spectra.

\subsection{The Swan bands}

The C$_{2}$ Swan bands  are detectable in all but the hottest RCB stars. They are 
not seen in either V3795\,Sgr or XX\,Cam with effective temperatures
of 8000\,K and 7250\,K, respectively. In our sample, they are first detectable in
VZ\,Sgr at $T_{\rm eff}$ = 7000\,K. The bands are very strong in the coolest
RCB stars like U\,Aqr and the HdC stars.
The leading bands of
the three sequences: $\Delta\nu$\,=\,+1, 0, and -1 are each considered. All bands have
blue-degraded bandheads. The (0,0) band of the
$^{12}$C$_2$ molecule  with its head at 5165\,\AA\ is
the strongest band of the entire Swan system.
The (1,0) and (0,1)  bandheads are at  4737\,\AA\ and 5636\,\AA, respectively.  All
three bands are
synthesized using detailed line lists including the blending atomic lines
and appropriate model atmospheres.
The (1,0), (0,0), and (0,1) $^{12}$C$_2$ bands are used to determine the C
abundance and, hence, to assess the carbon problem. 
The (0,1) band is generally a superior indicator of the C
abundance because it is less affected by blending atomic lines.
However, the (1,0) band is the focus of efforts to determine
the $^{12}$C/$^{13}$C ratio because the  $^{12}$C$^{13}$C bandhead
is shifted to 4745\,\AA\ and, thus, 8\,\AA\ clear of the blue-degraded $^{12}$C$_2$
band. For the (0,0) and (0,1)
bands, the $^{12}$C$^{13}$C lines are  mixed among the stronger $^{12}$C$_2$ lines.

Data required for synthesis of Swan bands include\,: wavelengths of the transitions,
excitation energies of the
lower levels, $gf$-values of the lines and the C$_2$ molecule's dissociation energy.
Accurate wavelengths for $^{12}$C$_2$ lines are taken from \citet{phillips68}.
Excitation energies are computed from the molecular constants
given by the latter
 reference.  The wavelength shift between a $^{12}$C$^{13}$C line and
the corresponding $^{12}$C$_2$ line is calculated using standard formulae for
the
vibrational and rotational shifts \citep{herzberg48, stawikowski64,
russo11}. Predictions for the bandhead wavelength shifts were
checked against the
measurements by \citet{pesic83}.

$gf$-values are calculated from the theoretical band oscillator strengths computed by
\citet{schmidt07}: $f$(1,0) $= 0.009414$, $f$(0,0) $=0.03069$, and $f$(0,1)
$=0.01015$. These theoretical
computations predict radiative lifetimes for the upper state of the Swan system that
are within a
few per cent of the accurate measurements by laser-induced fluoresence reported by 
\citet{naulin88}.
The C$_2$ dissociation energy is taken from an experiment involving multi-photon
dissociation of acetylene:
$D_0$(C$_{2}$) = 6.297 eV \citep{urdahl91}. Our molecular data for individual
$^{12}$C$_2$ lines --
$gf$-values and excitation energies -- are
in excellent agreement with values listed by \citet{asplund05} for their
determination of the solar C
abundance.  For detailed molecular line lists used in our  
analyses of C$_{2}$ bands, including the wavelengths, 
$J$-values of the lower level, 
the lower excitation potentials,
and the log $gf$-values, see Tables 2, 3, 4, 5, 6 and 7. 

\subsection{Atomic lines}

In order to ensure a satisfactory synthesis of a RCB spectrum, 
an accounting for the atomic lines at the wavelengths covered by the C$_2$ bands
is necessary,
most especially for the (1,0) 
$^{12}$C$^{13}$C bandhead which is
always weak and generally seriously blended. The region 4729-4748\,\AA\ 
was given especial attention. 
The procedure applied to the (1,0) band was followed for the
(0,0) and (0,1) bands.

Prospective atomic lines were first compiled
from the usual primary sources: the Kurucz database
\footnote{http://kurucz.harvard.edu}, the
NIST database\footnote{http://www.nist.gov},
the VALD database\footnote{http://vald.astro.univie.ac.at} and the
comprehensive multiplet table
for Fe\,{\sc i} \citep{nave94}.
Our next step was to identify the atomic lines in the spectrum of $\gamma$\,Cyg,
Arcturus, and the Sun
and to
invert their equivalent widths to obtain the product of a line's $gf$-value and the
element's abundance. For lines of a given species (e.g., Fe\,{\sc i}), the
assumption is that the
relative $gf$-values obtained from these sources may be applied  to a RCB
spectrum synthesis but an adjustment
may be needed to allow for an abundance difference between the source and the RCB.
After the adjustment for abundance differences between the sources, the $gf$-values
are  in agreement
within 0.1 dex (see Table\,8 for the individual estimates of the
$gf$-values as well as the adopted value).
For most lines the $gf$-values 
adopted are those 
derived from $\gamma$\,Cyg spectrum. For the 
lines which are not resolved in $\gamma$\,Cyg spectrum,
the $gf$-values are adopted from the solar spectrum.

Lines of C\,{\sc i} present in all RCB spectra are not present in the
reference spectra of $\gamma$\,Cyg, Arcturus and the Sun.
The C\,{\sc i} lines were identified using \citet{moore93}'s
multiplet table with
$gf$-values taken from the
NIST database. A C\,{\sc i} line is betrayed by the fact that a given
C\,{\sc i}
line has a similar strength in all RCB spectra.
In this regard the feature coincident with the 
$^{12}$C$^{13}$C (1,\,0) bandhead
is unlikely to be a very weak unidentified C\,{\sc i} line because
its strength varies from star to star. Note, for example, the absence or near
absence of this line in the spectra of V3795\,Sgr and V854\,Cen.
Furthermore, this line is stronger in the spectrum of $\gamma$\,Cyg,
where the C\,{\sc i} lines are very weak.

Initially, elemental abundances for RCB stars were
adopted from \citet{asplund00} and \citet{rao03}.
Then, equivalent widths  were measured off our spectra
and the abundances redetermined for RCB stars were 
found to be in good agreement with \citet{asplund00}.
In particular, we derived the Fe abundance from lines 
in the 4745-4810\,\AA\  window where C$_{2}$ contamination 
is minimal.
The Fe abundances derived from these Fe\,{\sc i} lines
are in good agreement with the Fe abundances
derived by \citet{asplund00} (see Table\,9).
These Fe abundances were adopted for deriving the   
$^{12}$C/$^{13}$C ratios in RCB stars. 
The uncertainties on the Fe abundance is used to 
derive the upper and lower limits to $^{12}$C/$^{13}$C ratios
in RCB stars (including U\,Aqr).
The metal abundances for the synthetic spectra are
adopted from \citet{asplund00} for most of the stars.
However, for V2552\,Oph we adopt the abundances from
\citet{rao03}.
We also assume the solar relative abundances 
with the correction of about +0.3 dex for the 
$\alpha$-elements at these metallicities, 
if these abundances are not measured in these stars.
Fe abundances are derived also for HdC stars and 
the cool RCB U\,Aqr.

\subsection{Spectrum synthesis of the C$_2$ bands}

For the spectrum synthesis,
we used the line-blanketed H-deficient model atmospheres by
\citet{1997A&A...318..521A}
and the UPPSALA spectrum synthesis BSYNRUN program.
For equivalent width analysis we used EQWRUN program.
The appropriate model atmosphere for a given RCB star was chosen using the
stellar parameters from \citet{asplund00}: effective temperature
$T_{\rm eff}$, surface gravity log \textit{g}, and microturbulence
$\xi_{\rm t}$.

The stellar parameters for the cool RCB star U\,Aqr and HdC stars are adopted from
\citet{1997A&A...318..521A} and \citet{garc09, garc10} and used with
the MARCS model
atmospheres \citep{gustafsson08} provided by
Kjell Eriksson (private communication)
used by \citet{garc09,garc10}.
For the four HdC stars and the cool RCB star U\,Aqr, we have derived the
microturbulence
($\xi_{\rm t}$) from Fe\,{\sc i} lines in the region of 4750-4960\,\AA\ since there are
no significant molecular bands in this wavelength region (Warner
1967).
The microturbulent velocity derived from Fe\,{\sc i} lines for U\,Aqr is
$\xi_{\rm t}$\,=\,5.0$\pm$2\,km\,s$^{-1}$ and the Fe abundance is
log\,$\epsilon$(Fe)\,=\,6.7$\pm$0.3, but adoption of a lower effective temperature,
$T_{\rm eff}$ = 5400 K, suggested by \citet{garc10} gives an Fe abundance of 6.5$\pm$0.3.
For HdC stars, the derived microturbulent velocities and Fe abundances are: for
HD\,137613,
$\xi_{\rm t}$\,=\,6.5$\pm$2\,km\,s$^{-1}$ and log\,$\epsilon$(Fe)\,=\,6.8$\pm$0.3, for
HD\,182040, $\xi_{\rm t}$\,=\,6.5$\pm$2\,km\,s$^{-1}$ and 
log\,$\epsilon$(Fe)\,=\,6.6$\pm$0.3, for
HD\,173409,
$\xi_{\rm t}$\,=\,6.0$\pm$2\,km\,s$^{-1}$ and log\,$\epsilon$(Fe)\,=\,6.6$\pm$0.3, and
for HD\,175893,
$\xi_{\rm t}$\,=\,6.0$\pm$2\,km\,s$^{-1}$ and
log\,$\epsilon$(Fe)\,=\,6.7$\pm$0.3. The other stellar parameters like $T_{\rm eff}$
and log \textit{g}, and
 the elemental abundances are judged from \citet{warner67},
\citet{1997A&A...318..521A}, and  \citet{garc09}.

 Stars with effective
temperature less than or about 7000\,K were selected for the analysis  of
their C$_{2}$ bands.
The C$_{2}$ molecular bands were synthesized
with the  line lists 
discussed above.
The synthesized spectrum was convolved with a Gaussian profile with a width that
represents the combined effect of stellar macroturbulence
and the instrumental profile. The synthesized spectrum is then matched to the
observed spectrum by adjustment of the
appropriate abundances.

\section{The carbon abundance}

If there were no carbon problem for C$_{2}$ bands, the $^{12}$C abundance 
derived by fitting each $^{12}$C$_2$
band would equal the input C abundance of the adopted model atmosphere to within the
margin implied by the  uncertainties arising from the errors assigned to the
model atmosphere parameters. (The changes in spectrum syntheses arising from
uncertainties in the basic data for the Swan bands and in the carbon isotopic ratio are
negligible.)

In Table\,10 the derived C abundances from C$_{2}$ bands for the RCB stars
are summarized for the three bands and for
models with C/He = 0.3, 1.0 and 3.0\%.  
Table\,11 similarly gives C abundances for the
four HdC stars and the cool RCB star U\,Aqr. 
In both Table\,10 and 11 we also give the C abundance from the C\,{\sc i}
lines but only for C/He=1\% models.
For a given model, the three bands give the same
C abundance to within 0.2 dex, a quantity comparable to the
fitting uncertainty. Along the sequence of models from C/He = 0.3 to
3.0\%, the derived C abundance decreases by about 0.2 dex for the warmest stars to
0.1 dex for the coolest stars or equivalently the carbon problem increases from the
warmest stars to the coolest stars. A carbon problem exists for all models
with C/He in the range from 0.3 to 3.0\%. Extrapolation of the 
C abundances in Table\,10 to
lower input C/He ratio suggests that elimination of the C problem
requires models with values of C/He across the range 0.3\% (VZ\,Sgr, R\,CrB) to
0.03\% (V2552\,Oph, SU\,Tau). Table\,11 suggests that C/He $\simeq$ 0.3\%  may
account for the HdC stars and cool RCB U\,Aqr. 
Adoption of $T_{\rm eff}$ = 6000K for U\,Aqr suggests C/He of 10\%, and not
in line with that of HdC stars. Hence, The $T_{\rm eff}$ = 5400K
is adopted for U\,Aqr over the $T_{\rm eff}$ = 6000K.
Discussion of this C/He range is postponed to
Section 5.

By way of illustrating the fits of the synthetic spectra to observed spectra
for the warm RCB stars, 
we show synthetic and observed spectra for SU\,Tau in Figures\,1,  and 2 for
the (0,0) and (0,1) C$_2$ bands, respectively. A corresponding figure for the
(1,0) figure is shown later.
The (0,1), (0,0) and
(1,0) bands each highlight a different issue.
For the HdC stars and the cool RCB U\,Aqr, the C$_2$ bands are very strong and the
issues
are somewhat different and related to the saturation of the lines.

For the (0,1) band,  a $^{12}$C
abundance is found to fit well the entire illustrated region except that right at the
bandhead the observed spectrum is shallower than that predicted.
This mismatch is not peculiar to SU\,Tau and is insensitive to the
choice of the C/He ratio.
This best fit for SU\,Tau demands a C abundance of 8.1 or, equivalently, 
presents a C
problem of 0.9 dex; the synthesis with a C abundance of 9.0 (i.e., zero C 
problem)
is obviously a very
poor representation of the observed spectrum.

Synthetic spectra for the (0,0) bands give results essentially identical to those
for the (0,1)  bands. The C abundance from the best-fitting synthesis as judged by the
fit to the C$_2$ lines away from the bandhead is within 0.2 dex of the values from the
(0,1) bands. The mismatch between synthesis and observation
at the bandhead is greater than for the (0,1) band
and extends over a greater wavelength interval than for the
(0,1) band.

A special difficulty occurs at the (1,0) $^{12}$C$_2$ bandhead because there are
strong atomic lines at and shortward of the bandhead. A line right at
the head is
a Fe\,{\sc i} line and those shortward of the head are C\,{\sc i} lines. These and
weaker atomic lines make it difficult to distinguish the C$_2$ contribution to
the spectrum from that of the atomic lines when the C$_2$ contribution is weak.

As long as the continuous opacity is provided by photoionization of neutral carbon,
the carbon problem (see Tables\,10 and 11)
raised by the C\,{\sc i} lines cannot be erased by changes to the
stellar parameters. 
The original carbon problem referred to the mismatch between the observed and
predicted strengths of C\,{\sc i} lines: the latter were stronger than the former by
an amount equivalent to about a 0.6 dex reduction in a line's $gf$-value. The
star-to-star
variation in this reduction across the RCB sample
was small: for example, the C\,{\sc i} problem for the
ten stars in Table\,10 spanned the small interval of $-0.3$ to $-0.9$ with a mean
value of $-0.7\pm0.1$ \citep{asplund00}.  This carbon problem's magnitude is
almost independent
of
the assumed C/He ratio for which the model is constructed, i.e., the difference
between the assumed and derived C abundance is maintained as C/He is adjusted.
The C\,{\sc i} lines included in present
syntheses
confirm the C problem. With $gf$-values from the NIST database, 
these lines demand a
gf-value decrease of  0.5 to 0.8 dex for the 
eleven stars in Table\,10 and the five stars in Table\,11.

\subsection{The $^{12}$C/$^{13}$C ratio}

The $^{12}$C$^{13}$C molecule's contribution to the spectra is assessed from the
(1,0) band. Unfortunately, there is an
unidentified atomic line coincident with the $^{12}$C$^{13}$C bandhead. Syntheses
show that this atomic line is a major contributor to the stellar 
feature in most
stars.
There is also strong atomic blending of the $^{12}$C$_2$ bandhead but the
$^{12}$C abundance
is provided  securely from the (0,0) and (0,1) bands.
Given these complications, our focus is on
determining whether the $^{12}$C/$^{13}$C ratio is close to the CN-cycle equilibrium
ratio (= 3.4), as
might be anticipated for a star produced by the FF scenario, or is a much higher
value, as
might be provided from the DD scenario. The intensity of a line from the
heteronuclear $^{12}$C$^{13}$C
molecule
and the corresponding line from the homonuclear $^{12}$C$_2$ molecule are 
related as $I(12-13)=
2I(12-12)/R$ where
$R$ is the $^{12}$C/$^{13}$C ratio.

Of particular concern to a determination of the $^{12}$C/$^{13}$C ratio is the
atomic line at
4744.39\,\AA\ which is coincident with the (1,0) $^{12}$C$^{13}$C bandhead. This line
is
present in the spectrum of $\gamma$\,Cyg, also  of the Sun and Arcturus. A line at this
wavelength is present in spectra of the hotter RCB stars  (V3795\,Sgr and XX\,Cam)
whose spectra show no sign
of the stronger (0,0)  C$_2$ band at 5165\,\AA.
The interfering line is
unidentified in \citet{hinkle00}'s
Arcturus atlas.  The line list given at the ccp7
website\footnote{http://ccp7.dur.ac.uk/ccp7/DATA/lines.bell.tar.Z}
 identifies the line as arising from a lower level in Fe\,{\sc i}  at 4.50 eV
but such a line and lower level
is not listed by \citet{nave94} in their comprehensive study of the Fe\,{\sc i} 
spectrum.
The line list given in ccp7 is
from \citet{bell89}, an unpublished line list.
Although this line  is assigned in Table\,8 to this (fictitious?) Fe\,{\sc i}
transition, the
lack of a positive
identification is not a serious issue except, as we note below, perhaps for the
minority RCB stars.
Given that the $gf$-value of the line is fixed from the line's strength in
spectra of stars that span the temperature range of the RCBs ($\gamma$\,Cyg,
Arcturus, and the Sun),
alternative identifications have little effect on the predicted strength of the line in
a RCB or a HdC star. We assume it is a Fe\,{\sc i} line and 
predict its strength from
the inferred $gf$-value (Table\,8) and the Fe abundance derived from a sample of
Fe\,{\sc i} line in the same region (see above). Table\,9 lists our
derived Fe abundance, the Fe abundance from \citet{asplund00}, and the Fe
abundance obtained on the assumption that the entire $^{12}$C$^{13}$C 
bandhead is
attributable to the Fe\,{\sc i} line. 
There is good agreement between our Fe abundance and that
derived from different spectra by \citet{asplund00}. Perfect agreement
would not be expected for several reasons: for example, the stars are somewhat
variable even out of decline and our spectra are not those analysed
by Asplund et al. (2000).
The difference between the mean Fe abundance and the
abundance 
required to fit the feature at the $^{12}$C$^{13}$C bandhead is a rough measure of the
inferred  molecular contribution to the feature.

Stars are discussed in the order of decreasing effective temperature.
For all the stars synthetic spectra are computed for a
model with the parameters given in Table\,12 and
with C/He= 1.0\%.
The $^{12}$C$_2$ bands are fitted and then several syntheses are
computed for various values of the isotopic ratio $R$.
The estimates of $^{12}$C/$^{13}$C ratio are  given in Table\,12.

{\bf VZ\,Sgr:} Observed and synthetic spectra around the (1,0) band are shown in
Figure\,3 for this minority RCB star.
At 4745\,\AA, the atomic line
 (here assumed to be the Fe\,{\sc i} line from Table\,8) is too weak to
account for the observed feature; Table\,9 shows that the Fe
abundance must be increased by about 1 dex to remove the necessity for a
contribution from $^{12}$C$^{13}$C.
A contribution from $^{12}$C$^{13}$C seems
necessary with
$R\simeq 3-6$,
 a value suggestive of CN-cycling.  
The observed $^{12}$C$^{13}$C band is very weak, and taking 
into account the signal-to-noise ratio, the expected band asymmetry 
is not evident. The blending Fe\,{\sc i} line at the $^{12}$C$^{13}$C
 bandhead further removes the expected asymmetry. However, 
the blending Fe\,{\sc i} line is very weak, the residual spectrum, 
observed/synthesis\,(pure C$_2$ with R = 4),
suggests the presence of the contaminating line at 4744.39\,\AA\
within the uncertainties.

Since the relative metal abundances
for VZ\,Sgr, a minority RCB, are
non-solar \citep{lambert94},
the identity of the 4745\,\AA\ atomic line may affect the conclusion that
this line is an unimportant contributor to the molecular bandhead. For example
,
VZ\,Sgr is a
minority RCB
especially rich in Si and S  ([Si/Fe] $\sim$ [S/Fe] $\sim$ 2)
and a blending line from these elements may reduce the
need for
a $^{12}$C$^{13}$C
contribution. However, a search of multiplet tables of Si\,{\sc i} \citep{martin83}
and S\,{\sc i} \citep{martin90, kaufman93} did not uncover an
unwanted blend. Thus, we suppose that VZ\,Sgr is rich in $^{13}$C.

{\bf UX\,Ant:} There is a strong (1,0) $^{12}$C$_2$  contribution to the spectrum.
The predicted profile of the bandhead is broader than the observed head which
is distorted by very strong cosmic ray hits on the raw frame.
The  Fe\,{\sc i} line is predicted to be a weak contributor to the
feature at the $^{12}$C$^{13}$C wavelength.
Values of $R$ in the range 14 to 20 fit the observed feature quite
clearly, a synthesis with
$R=3.4$ provides a bandhead that is incompatible with the observed head
(Figure\,4).

{\bf RS\,Tel:} Observed and synthetic spectra  shown in Figure\,4 indicate that
the
Fe\,{\sc i} line at the $^{12}$C$^{13}$C band head accounts well for the observed
feature and
thus $R > 60$ is all that can be said for the carbon isotopic ratio from this
spectrum of relatively low S/N ratio.

{\bf R\,CrB:} Figure\,5 shows observed and synthetic spectra.
The Fe\,{\sc i} line at the $^{12}$C$^{13}$C bandhead accounts  for the
observed feature.
Given that the identity of the line's carrier is uncertain, a conservative view must
be that
$^{12}$C$^{13}$C contributes  negligibly to the observed feature and $R > 40$ is
estimated . It is
clear, however,
that $R=3.4$ is excluded as a possible fit.

{\bf V2552\,Oph:} The spectrum of this recently discovered RCB is very similar to that
of R\,CrB \citep{rao03} but for its stronger N\,{\sc i} lines and weaker
C$_2$ bands
(Figure\,1 of \citet{rao03}). The $^{12}$C$_2$ bandhead is very largely
obscured by the
overlying Fe\,{\sc i} line.
The apparent   $^{12}$C$^{13}$C bandhead is
almost entirely reproduced
by the atomic line. A high $R$ value cannot be rejected but $R=3.4$ may be excluded
(Figure\,5). $R > 8$ is our estimate.

{\bf V854\,Cen:} This RCB with low metal abundances provides a clean spectrum in the
the region of the (1,0) Swan bands (Figure\,6). The $^{12}$C$_2$ head is well
fitted with a synthetic spectrum. Very high S/N ratio  spectra are
necessary to set strict limits on the $^{12}$C$^{13}$C bandhead but it is
clear that the blending Fe\,{\sc i} line is a weak contributor; the Fe 
abundance must be increased by 1.5 dex to eliminate the need for a 
$^{12}$C$^{13}$C contribution.
A ratio $R=3.4$ is
firmly excluded. Values of $R$ in the range 16 to 24 are suggested.

{\bf V482\,Cyg:} The Fe\,{\sc i} line accounts well for the 
observed feature (Figure\,6) with
a lower limit for the isotopic ratio $R>100$.

{\bf SU\,Tau:} 
At the $^{12}$C$^{13}$C bandhead, the atomic line makes a
dominant contribution  but the profile of the observed feature suggests that
the Swan band is contributing to the blue of the atomic line (Figure\,7):
$R$ seems to be $>$24.
The $R=3.4$ synthetic
spectrum is clearly rejected as a fit to the observed spectrum.

{\bf V\,CrA:} The Fe\,{\sc i} line at the $^{12}$C$^{13}$C bandhead,
and the expected band asymmetry, is seemingly quite
unimportant but V\,CrA is another minority RCB so that the identity of the line's
carrier may be relevant here (see above notes on VZ\,Sgr). The $^{12}$C$_2$ band is
quite
strong (Figure\,7). With the blending line assigned to Fe\,{\sc i}, the observed
$^{12}$C$^{13}$C  bandhead is well fit with $R\simeq$8 to 10.
Our derived $^{12}$C/$^{13}$C ratio is in agreement with the upper limit of
the range set by \citet{rao08} from the same spectrum. Note that an
additional line about 0.6\,\AA\ to the blue of the $^{12}$C$^{13}$C  bandhead
 is seen in this spectrum.

{\bf GU\,Sgr:}
Presence of the $^{12}$C$^{13}$C band
is doubtful because atomic lines may account fully for the bandhead and the region 
just to
the
blue: $R$ seems to be in the range $>$40 (Figure\,8).

{\bf FH\,Sct:} Spectrum synthesis shows that $^{12}$C$_{2}$ makes a minor contribution
to the
observed spectrum (Figure\,8) but the $^{12}$C abundance may be established from
the (0,0) and (0,1) bands.   The ratio $R > 14$ may be set and the CN-cycle's
limit of R=3.4 is excluded.

{\bf U\,Aqr:}  The (1,0) $^{12}$C$_2$
band
is so strong (Figure\,9) that
the uncertainty over $R$  is
dominated
by the derivation of the $^{12}$C abundance from the very saturated  (1,0) $^{12}$C$_2$
lines.
 The carbon abundance  from (the also saturated)
(0,1) C$_2$ band is used  with the (1,0) $^{12}$C$^{13}$C blend to derive the
 $^{12}$C/$^{13}$C ratio. 
The Fe abundance is derived from several lines longward of the (1,0) $^{12}$C$
^{13}$C bandhead.
A $^{12}$C/$^{13}$C ratio in the range 110 to 120 is obtained.  

{\bf HdC stars:}  Syntheses of the (1,0) band are shown in Figures\,10\,-\,11 for
the four
HdC stars with the $^{12}$C abundance set in each case by the fit to the
(0,1) band (see Figure\,12 for a typical fit).   
In contrast to U\,Aqr, the $^{12}$C$^{13}$C bandhead
is well fit by the blending atomic lines with the Fe abundance obtained from
lines longward of the bandhead.
The derived $^{12}$C/$^{13}$C ratio is
$>$100 for HD\,137613,
$>$400 for HD\,182040, $>$100 for HD\,175893, and
$>$60 for HD\,173409.

\section{Discussion - C$_2$ and the Carbon Problem}

The carbon problem as it appears from
the analysis of C\,{\sc i} lines is discussed fully
by \citet{asplund00}. In brief, when analysed with state-of-the-art
H-poor model atmospheres \citep{1997A&A...318..521A} constructed for a C/He ratio
($ = 1$\%) representative of EHe stars where direct determinations of C and He
abundances are possible, the C\,{\sc i} lines return a C abundance that is about
0.6 dex less than the input abundance of log $\epsilon$(C) = 9.5.
The derived  abundance varies little from star-to-star: 13
of the 17 analysed RCBs have abundances between 8.8 and 9.0 and the mean from the
set of 17 is 8.9$\pm0.2$. A similar result is apparent from Tables\,10 and 11
where the C abundance from C\,{\sc i} lines from our spectrum syntheses is 
quoted.
The discrepancy of 0.6 dex between assumed and
derived C abundance is the (C\,{\sc i}) carbon problem.
As the C/He ratio of a model atmosphere is adjusted, the carbon problem
(i.e., the 0.6 dex difference between assumed and derived C abundances) 
persists
until a low C/He reached. This persistence arises because the
continuous opacity arises from photoionization of neutral carbon from excited
levels.

Tables\,10 and 11 also show that the C$_2$ bands exhibit a carbon problem but one that
differs from that shown by the C\,{\sc i} lines in several ways: 
(i) the C abundance from
C$_2$ bands is almost independent of the assumed C/He ratio unlike the 
abundance
from C\,{\sc i} lines; (ii) the star-to-star spread in C abundances from C$_2$
 bands
is larger than found from the C\,{\sc i} lines; (iii) the C abundance from C$_
2$ bands 
is somewhat more sensitive than that from C\,{\sc i} lines to changes in the 
adopted
atmospheric parameters as reflected by Table\,13 where models spanning the 
effective temperature and surface gravity uncertainties suggested by Asplund et al.
(2000)
are considered.

Taken in complete isolation, inspection of Table\,10 suggests that a C/He ratio
of less than 0.3\% can be found for which the ratio adopted in the construction of
the model atmosphere is equal to that derived from the C$_2$ bands. For the RCB
stars, Table\,10 suggests C/He ratios running from
about 0.3\% for VZ\,Sgr and R\,CrB down to 0.03\% for GU\,Sgr and FH\,Sct. For the
HdC stars (Table\,11), a C/He of slightly larger than 0.1\% is suggested. 
However, \citet{asplund00} remark that C/He $\leq$ 0.05\% is required to eliminate the
C\,{\sc i} carbon problem by lowering the carbon abundance to the point that
photoionization of neutral carbon no longer is the dominant opacity source.

Resolution of the carbon problems by invoking low C/He ratios
deserves to be tested fully
by constructing model atmospheres with lower C/He ratios and appropriate
abundances for other elements and determining the
C abundances from C\,{\sc i} and C$_2$ lines. \citet{asplund00} recognized this
possible way to address the C\,{\sc i} carbon problem but discounted it on several
grounds: (i) removal of carbon photoionization  as the dominant continuous opacity
makes it difficult to account for the near-uniformity of the C\,{\sc i} equivalent
widths across the RCB sample, especially as O abundance varies from
star-to-star; (ii) an inverse carbon problem is created for the
C\,{\sc ii} lines at 6578\,\AA\ and 6582\,\AA\ which are seen in the hottest RCBs;
and (iii) these low C/He ratios for RCB stars
are at odds with the higher  ratios obtained directly from
He and C lines for EHe stars which one assumes are  intimate relatives of the
RCB and HdC stars. \citet{asplund00} noted that published analyses of EHe stars
gave the mean C/He = 0.8$\pm$0.3\% over a range 0.3-1.0\% with three unusual EHe stars
providing  much lower ratios (0.002\% to 0.2\%). \citet{2006ApJ...638..454P} confirmed
the C/He ratios for the leading group of EHe stars. 
From the following references 
\citep{pandey01, 2006ApJ...638..454P, pandey11, 
jeffery98, drilling98,2006MNRAS.369.1677P, 
jeffery93, jeffery99, jeffery97},
that also includes the recent analyses of these EHe stars, 
the mean value of C/He = 0.6$\pm$0.3\% is noted.

The RCB-EHe mismatch of their
C/He ratios invites two responses: (i) the carbon problems for the RCB stars should
be resolved on the assumption that their C/He ratios and those
of the EHe stars span  similar ranges; or, (ii) as a result of different evolutionary
paths, the C/He ratios of RCB and EHe stars span different ranges.

After considering a suite of possible explanations for their carbon
problem, \citet{asplund00} proposed that the actual atmospheres of
the RCB stars differed from the theoretical atmospheres in that the
temperature gradient was flatter than predicted. Hand-crafted atmospheres
were shown to solve the C\,{\sc i} carbon problem. However, the issue of accounting
for the additional heating and cooling of the hand-crafted atmospheres was
left unresolved. In principle, the change in the temperature structure -- a heating
at modest optical depths -- will require a higher C abundance to account for the
C$_2$ bands so that the C\,{\sc i} and C$_2$ carbon problems might be
both eliminated.

Further exploration of the carbon problem was pursued by \citet{pandey04}
who observed the 8727\,\AA\ and 9850\,\AA\ [C\,{\sc i}] lines in a sample of
RCB stars. The 8727\,\AA\ line gives a more severe carbon problem than the
C\,{\sc i} lines, say 1.2 dex versus 0.6 dex for C/He=1.0\% model atmospheres. 
In part, this difference
might be reduced by a revision of the effective temperature scale because
the [C\,{\sc i}] line being of low excitation potential has a temperature
dependence relative to the continuous carbon opacity from highly
excited levels. The 9850\,\AA\
line may give a similar carbon problem to the C\,{\sc i} lines or the
forbidden line may be blended with an unidentified line.
To account for the 8727\,\AA\ carbon problem, \citet{pandey04}
considered introducing a chromospheric temperature rise to the theoretical
model photospheres. Such a chromosphere with LTE produces emission at the
C$_2$ bandheads and offers a qualitative explanation for the fact that the
best-fitting synthetic spectra from the theoretical photospheres (i.e., no
chromospheric temperature rise) are deeper than the observed spectra at the
bandheads.

Analysis of the C$_2$ bands suggests a novel clue to the C$_2$ carbon problem.
As Figure\,13 shows the C abundance from the C$_2$ bands is correlated
with the O abundance derived by \citet{asplund00} from O\,{\sc i} and/or
[O\,{\sc i}] lines. The points are distributed about the  relation
C/O $\sim 1$.
(The C abundance from the C\,{\sc i} lines is not
well correlated with the O abundance and most points fall below the
C/O $= 1$ locus).
In Figure\,14, the EHe stars are plotted 
along with RCB stars. The RCB stars may connect those EHe stars of 
very low C/He with the majority of higher C/He.

Perhaps a more powerful clue is the fact that the Fe abundance of the RCB and
HdC stars is uniformly sub-solar. The mean Fe abundance excluding the minority
RCBs is 6.5 or 1.0 dex less than the solar Fe abundance. EHe stars show a similar
spread
and mean Fe abundance of 6.7 \citep{jeffery11}.

\section{Discussion - The $^{12}$C/$^{13}$C ratio and the origin of the
RCBs}

Discussion of the determinations of the $^{12}$C/$^{13}$C ratio may be
focussed on three main points.

First, the ratio is low in the two minority RCBs VZ\,Sgr and V\,CrA. Unless there is
an unidentified line from an element with an overabundance in a minority RCB
star, VZ\,Sgr has a $^{12}$C/$^{13}$C ratio equal within the measurement
uncertainty to the equilibrium ratio for the H-burning CN-cycle. The ratio is
higher  ($\simeq 8$) for  V\,CrA  but considerably lower than the upper limits
set for majority RCBs. (\citet{rao08} gave the ratio as 3.4 for V\,CrA
but apparently did not include the factor of two arising from the fact that
$^{12}$C$^{13}$C is not a homonuclear molecule, i.e., 3.4 should be 6.8.)
In some respects, V854\,Cen is a minority RCB star and its $^{12}$C/$^{13}$C ratio
of 18 is also generally lower than representative upper limits for majority
RCB stars.

Second, the $^{12}$C/$^{13}$C ratio for all majority RCBs is much larger than
the equilibrium ratio for the CN-cycle.\,\footnote {We note that \citet{goswami10}'s
observation that the ratio is about 3.4 for U\,Aqr is incompatible with our
spectra by
simple inspection of Figure\,9.}

Third, there appears to be a range in the $^{12}$C/$^{13}$C ratios
among majority RCBs. The star with the lowest ratio appears to be UX\,Ant
(Figure\,4) for which the predicted blend of atomic lines at the (1,0)
$^{12}$C$^{13}$C bandhead accounts for less than half of the strength of the
observed absorption feature.
Similarly for U\,Aqr, the atomic lines at the bandhead account for about half
of the observed feature but because the C$_2$ lines are strong the isotopic
bandhead translates to the ratio $^{12}$C/$^{13}$C $\simeq 110$.
For these cases at least, it seems likely that the $^{12}$C$^{13}$C bandhead 
is present in our spectra.
Within the uncertainties associated with the
blend of atomic lines, other RCBs yield a lower limit to the isotopic
ratio.  This limit is highest for the HdCs where the C$_2$ bands are very
strong. Note how the atomic lines account remarkably well for the observed 
spectrum between the $^{12}$C$_2$ and $^{12}$C$^{13}$C (1,0) 
bandheads.

Our results are in good agreement with published results for the
few stars previously analysed (Table\,12). In the case of R\,CrB,
\citet{cottrell82}
 determined a lower limit of 40 from the C$_2$ (0,1) band.
\citet{fujita77} from spectra of the CN Red system set lower limits of 500 for
HD\,137613 and $>$100 for HD\,182040. The latter limit was also reported by
\citet{climenhaga60} from high-resolution photographic spectra of C$_2$ bands.
Lower limits set by \citet{garc09, garc10} are not
competitive with those from CN or our limits from C$_2$.

Prospects for a low $^{12}$C/$^{13}$C ratio in the atmosphere of a
product of the DD scenario are dim. In the scenario's cold version (i.e., no
nucleosynthesis as a result of the merger), the C is provided by the
He shell of the C-O white dwarf and quite likely also by layers of the
C-O core immediately below the He shell.  The latter contribution will
be devoid of $^{13}$C. In the He shell, $^{13}$C may be present in the
outermost layers as a result of penetration of protons from the H-rich
envelope of the former AGB star into the He shell. Very slow penetration
results in the build up of a $^{12}$C/$^{13}$C ratio of about three
and inhibits somewhat the conversion of the $^{13}$C to $^{14}$N, as usually
occurs in the CN-cycle. This mechanism sustains the favored neutron
source for the $s$-process in AGB stars; the $^{13}$C($\alpha,n)^{16}$O
is the neutron source.
The He is provided almost
entirely by the He white dwarf which will have very little carbon
but abundant nitrogen as a result of H-burning by the CNO-cycles.

A cold merger of the He white dwarf with the He shell of the C-O
white dwarf results in a C/He ratio (see \citet{pandey11}, eqn. 1)

\begin{equation}
\frac{{\rm C}}{{\rm He}}\simeq\frac{{\rm A(He)}}{{\rm A(C)}}\frac{\mu({\rm C})_{{\rm
C-O:He}}M({\rm C-O:He)}}{M({\rm He)}}
\end{equation}

where $\mu$(C)$_{C-O:He}$ is the mass fraction
of $^{12}$C in the He shell, $M$(C-O:He) is the mass of the He shell
and $M$(He) is the mass of the He white dwarf. With plausible values for
the quantities on the right-hand side of the equation, i.e.,
$\mu$(C)$_{C-O:He}$ $\simeq 0.2$, $M$(C-O:He) $\simeq 0.02M_\odot$, and
$M$(He) $\simeq 0.3M_\odot$, one obtains C/He $\simeq 0.4$\%, a value at the
lower end of the C/He range found for EHe stars. Additional $^{12}$C
is likely provided by mixing with the layers of the C-O white dwarf
immediately below the He shell.

The attendant $^{12}$C/$^{13}$C ratio will be a maximum if these
latter contributions are absent. Then, this ratio will depend on the
fraction of the He shell over which $^{13}$C is abundant (relative to
$^{12}$C), say $^{12}$C/$^{13}$C $\sim$ $3/f_{13}$ where $f_{13}$ is the
fraction of the He shell which is rich in $^{13}$C.
For
$f_{13}$ $\sim 0.1$ and $0.01$, the predicted isotopic ratio is
30 and 300, respectively. These estimates must be increased when the
mixing at the merger includes the layers of the C-O white dwarf immediately
below the He shell.

At present, there are no reliable {\it ab initio} calculations of the
mass of the $^{13}$C-rich layer in the He shell of an AGB star. Additionally,
one is making a bold assumption that the He shell of the C-O white dwarf which
accepts merger with the He white dwarf resembles the He shell of a AGB star.
Mass estimates relevant to an AGB star
may be obtained by the fit to observed $s$-process abundances. 
\citet{gallino98} 
 suppose that the $^{13}$C rich layer amounts to about 1/20 of the mass
in a typical thermal pulse occurring in the He shell. 
With $f_{13} \simeq 0.05$,
the isotopic ratio is 60. 
Interestingly, this is
consistent with our inferred range.
This will increase as the
C-O white dwarf contributes to the mixing. Destruction of $^{13}$C by
$\alpha$ particles may
occur in a `hot' phase during the merger and thus also raise the isotopic
ratio; the $^{13}$C($\alpha,n)$ destruction rate is roughly a factor of 100 
faster than
the $^{14}$N$(\alpha,\gamma$) rate providing $^{18}$O.
The low isotopic ratio for minority RCB stars -- VZ\,Sgr, and V\,CrA -- are
unexplained as are their distinctive elemental abundances.

\section{Concluding remarks}

The C$_{2}$ Swan bands are present in spectra of the HdC stars
and in all but the hottest RCB stars. Analysis of these bands provides an
alternative to the C\,{\sc i} lines of providing estimates of the C abundance
and in addition provide an opportunity to estimate the $^{12}$C/$^{13}$C
ratio. When analysed with Uppsala model atmospheres,  the C$_{2}$ bands return a
C abundance that is almost independent of the C/He ratio assumed in construction of
the model atmosphere. If consistency between assumed and derived C abundances is
demanded, the C$_{2}$ bands imply C/He ratios across the RCB and HdC sample run in the
range  0.03\% to 0.3\%, a range that is notably lower than the range 0.3--1.0\%
found
from a majority of EHes. This mismatch, if not a reflection of different modes of
formation, implies that the C abundances for RCB and HdC stars are subject to a
systematic error. Therefore, it appears that a version of the carbon problem
affecting the C\,{\sc i} lines \citep{asplund00} applies to the C$_{2}$ lines.
Although alternative explanations can not yet be totally eliminated, it appears
that higher order methods of model atmosphere construction are needed in order to
check that Asplund et al.'s suggestion of the real atmospheres have a flatter
temperature gradient than predicted by present state-of-the-art model
atmospheres. Nonetheless, that the carbon abundances 
derived from C$_{2}$ Swan bands 
are the real measure of the carbon abundances in these stars
cannot be ruled out.

There is evidence for the presence of detectable amounts of $^{13}$C in the
spectra of a few RCB stars and especially for the minority RCB stars. For the
other RCBs and the HdC stars, lower limits are set on the carbon isotopic
ratio. Apart from the minority RCB stars, the estimates of the carbon isotopic
ratio are consistent with simple predictions for a cold merger of a He white
dwarf with a C-O white dwarf.

The minority RCB stars are an enigma. Their distinctive pattern of elemental
abundances remains unaccounted for; for example, V\,CrA has a Fe deficiency of
2 dex but [Si/Fe] $\sim$ [S/Fe] $\sim$ [Sc/Fe] $\simeq$ 2 with [Na/Fe]
$\sim$ [Mg/Fe] $\sim$ 1 (Rao \& Lambert 2008). Add to these anomalies, the
$^{12}$C/$^{13}$C ratio is much lower than found for the majority RCB stars
unless the (1,0) Swan $^{12}$C$^{13}$C bandhead is blended with an as yet
unidentified atomic line whose strength is unsuspected from examination of
the spectra of majority RCB stars. The enigma calls for additional
observational insights.

\acknowledgments
We thank Kjell Eriksson and Anibal Garc{\'{\i}}a-Hern{\'a}ndez for the 
H-deficient models of the cool stars, particularly Kjell Eriksson
for providing these models in our required format.
We thank Kameswara Rao for providing us the 
spectra of V854\,Cen and V\,CrA.
We also thank Nils Ryde for making available the  C$_{2}$ line list
by Querci et al. (1971) as a sample.  
DLL wishes to thank the Robert A. Welch Foundation of Houston, Texas for
support through grant F-634.

\clearpage
\begin{figure}
\epsscale{1.0}
\plotone{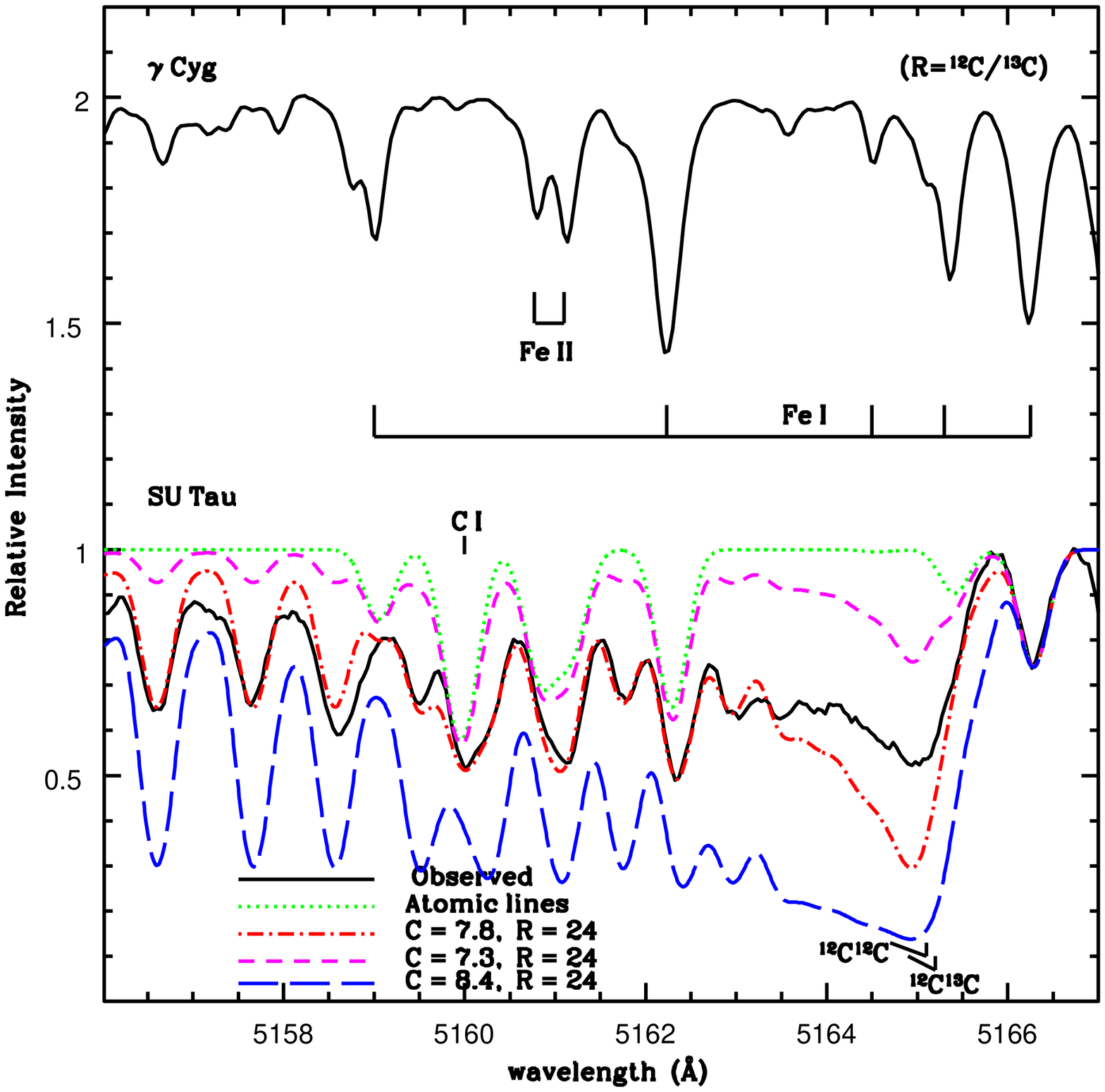}
\caption{Observed and synthetic spectra of the (0,\,0) C$_{2}$ band for
SU\,Tau. Synthetic spectra are plotted for different values of the C abundance -- see key on the figure. The spectrum of the $\gamma$\,Cyg is
plotted with the positions of the key lines marked.
\label{}}
\end{figure}

\clearpage
\begin{figure}
\epsscale{1.0}
\plotone{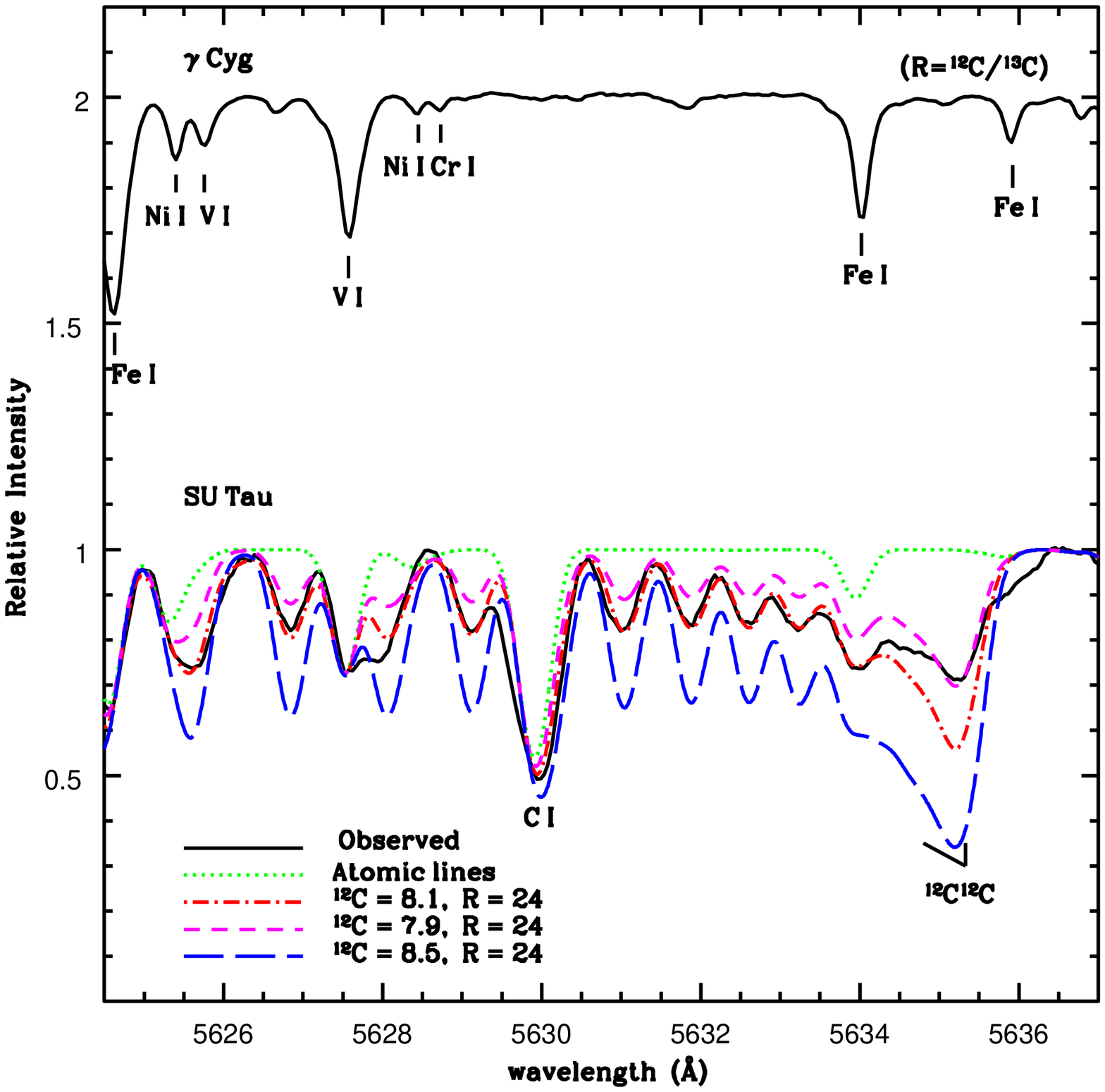}
\caption{Observed and synthetic spectra of the (0,\,1) C$_{2}$ band for
SU\,Tau. Synthetic spectra are plotted for different values of the C abundance -- see key on the figure. The spectrum of the $\gamma$\,Cyg is
plotted with the positions of the key lines marked.
\label{}}
\end{figure}

\clearpage
\begin{figure}
\epsscale{1.0}
\plotone{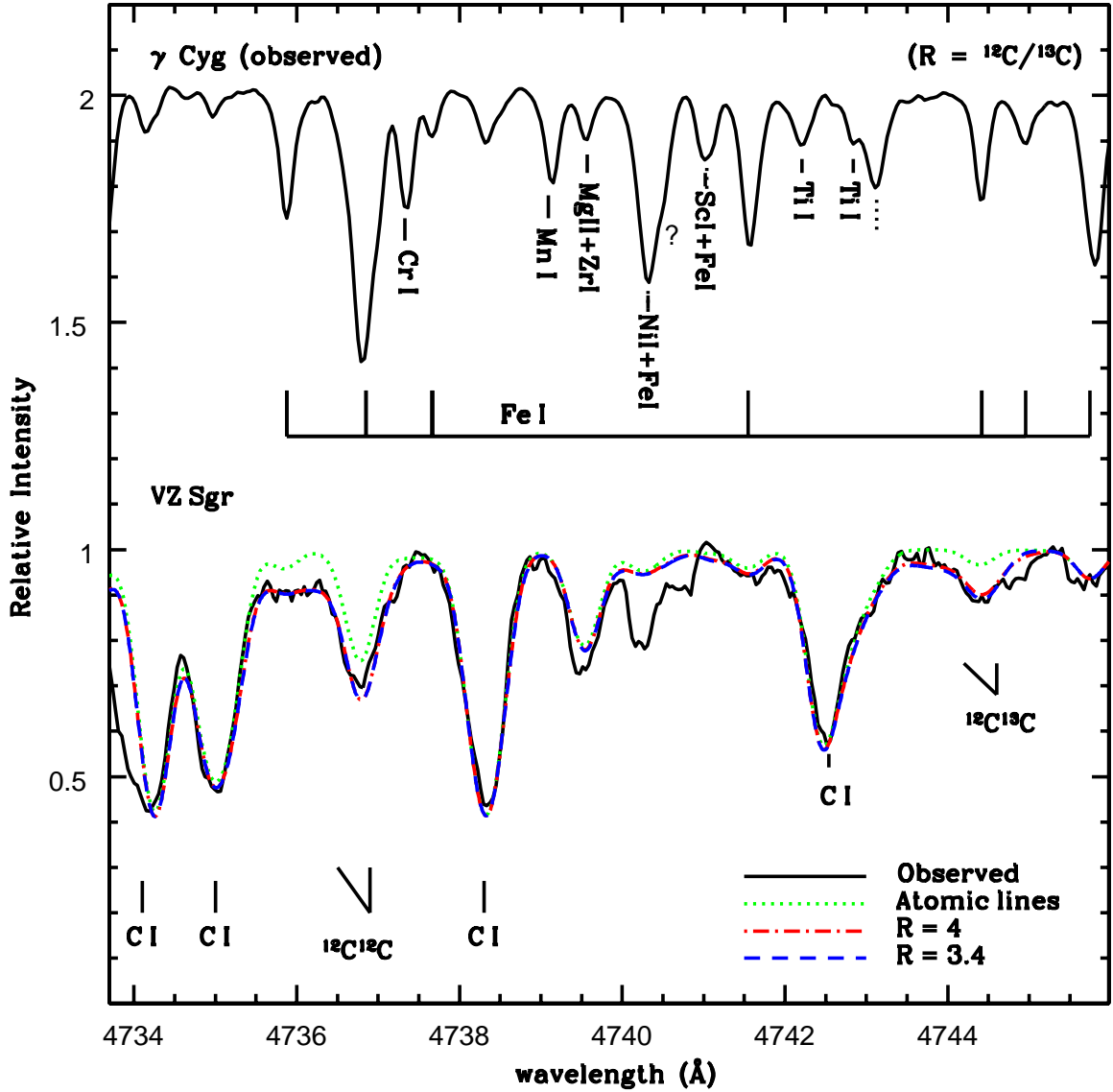}
\caption{Observed and synthetic spectra of the (1,0) C$_{2}$ bands for
VZ\,Sgr. Synthetic spectra are plotted for the values
of the isotopic ratios (R) shown in the keys and for a spectrum
with just the atomic lines.
The spectrum of $\gamma$\,Cyg is also plotted -- the positions of the
key lines are also marked -- the dotted line represents the blending
of the one or more atomic lines.
\label{}}
\end{figure}

\begin{figure}
\epsscale{1.0}
\plotone{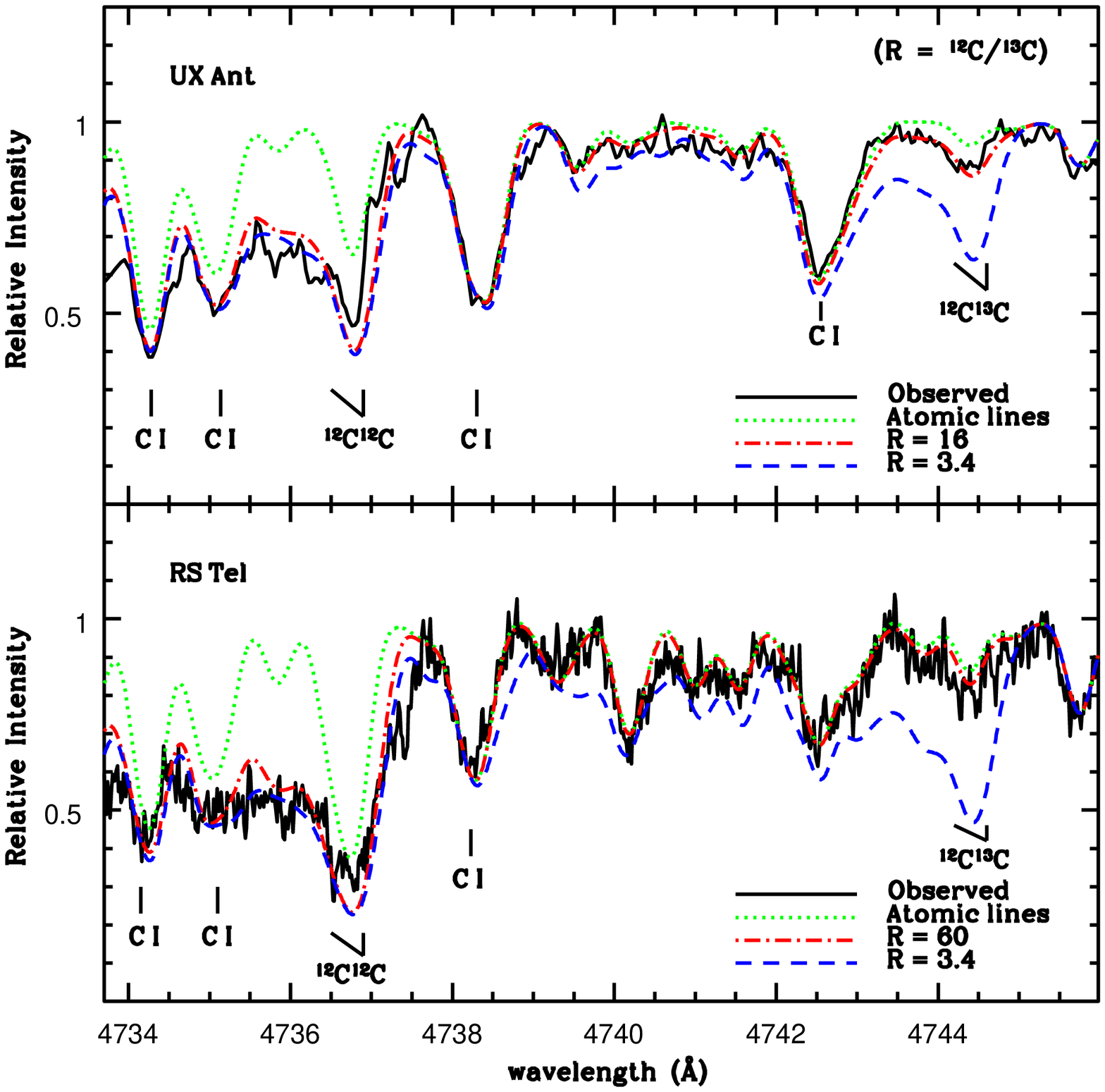}
\caption{Observed and synthetic spectra of the (1,0) C$_{2}$ bands for
UX\,Ant and RS\,Tel. Synthetic spectra are plotted for the values
of the isotopic ratios (R) shown in the keys and for a spectrum
with just the atomic lines. The positions of the key lines are also marked.
\label{}}
\end{figure}

\begin{figure}
\epsscale{1.00}
\plotone{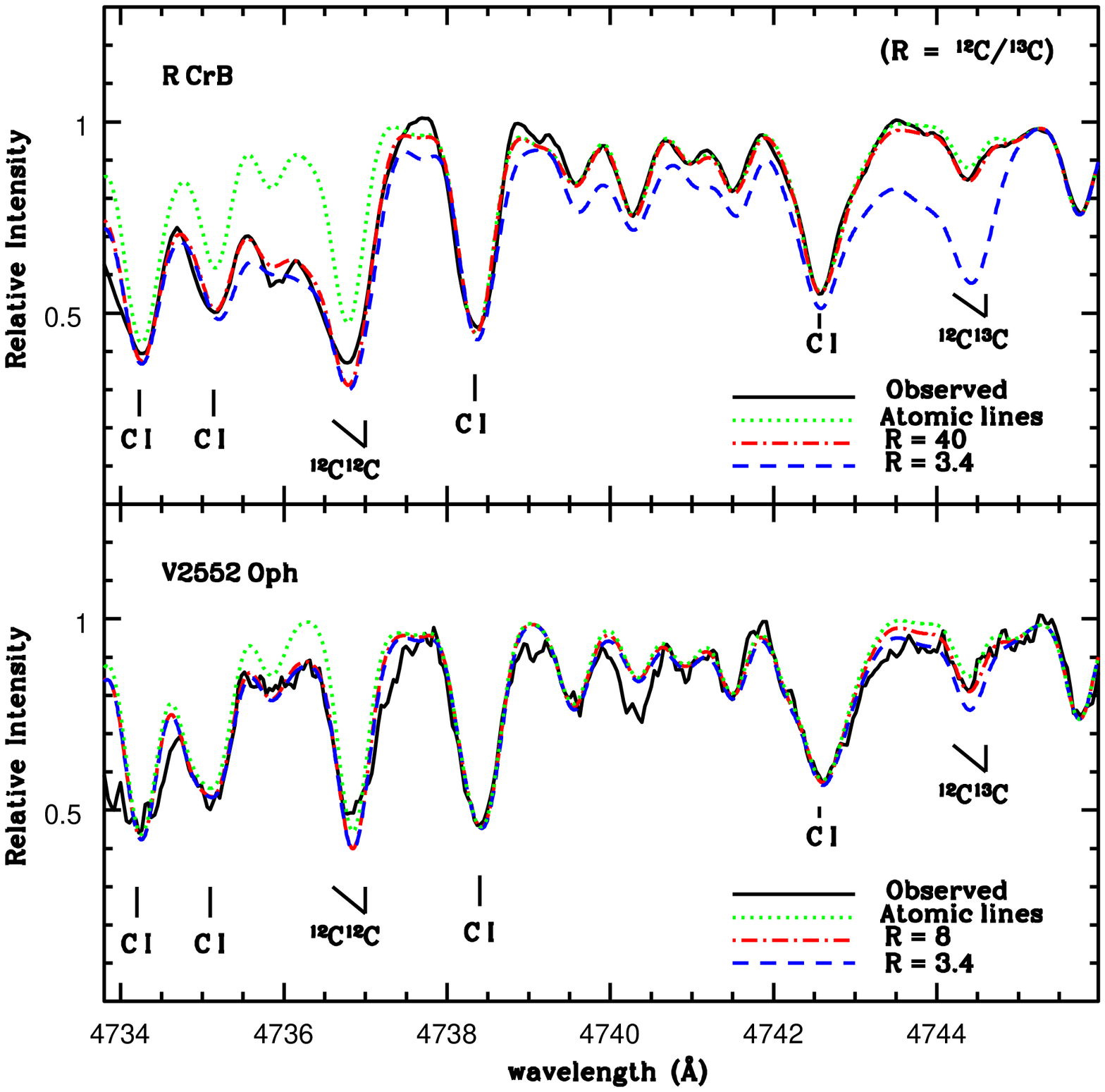}
\caption{Observed and synthetic spectra of the (1,0) C$_{2}$ bands for
R\,CrB and V2552\,Oph. Synthetic spectra are plotted for the values
of the isotopic ratios (R) shown in the keys and for a spectrum
with just the atomic lines.  The positions of the key lines are also marked.
\label{}}
\end{figure}

\begin{figure}
\epsscale{1.0}
\plotone{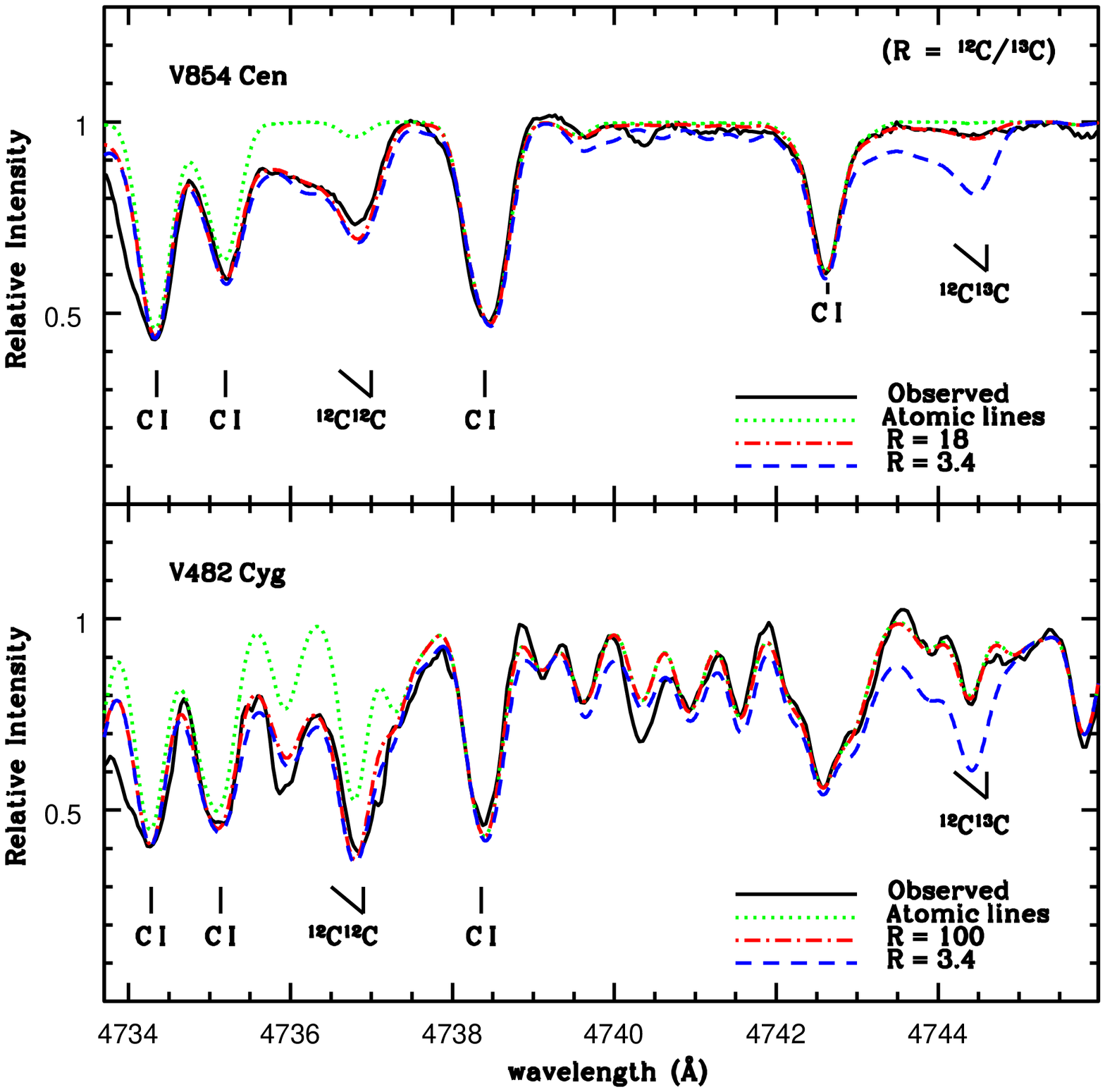}
\caption{Observed and synthetic spectra of the (1,0) C$_{2}$ bands for
V854\,Cen and V482\,Cyg. Synthetic spectra are plotted for the values
of the isotopic ratios (R) shown in the keys and for a spectrum
with just the atomic lines.  The positions of the key lines are also marked.
\label{}}
\end{figure}

\begin{figure}
\epsscale{1.0}
\plotone{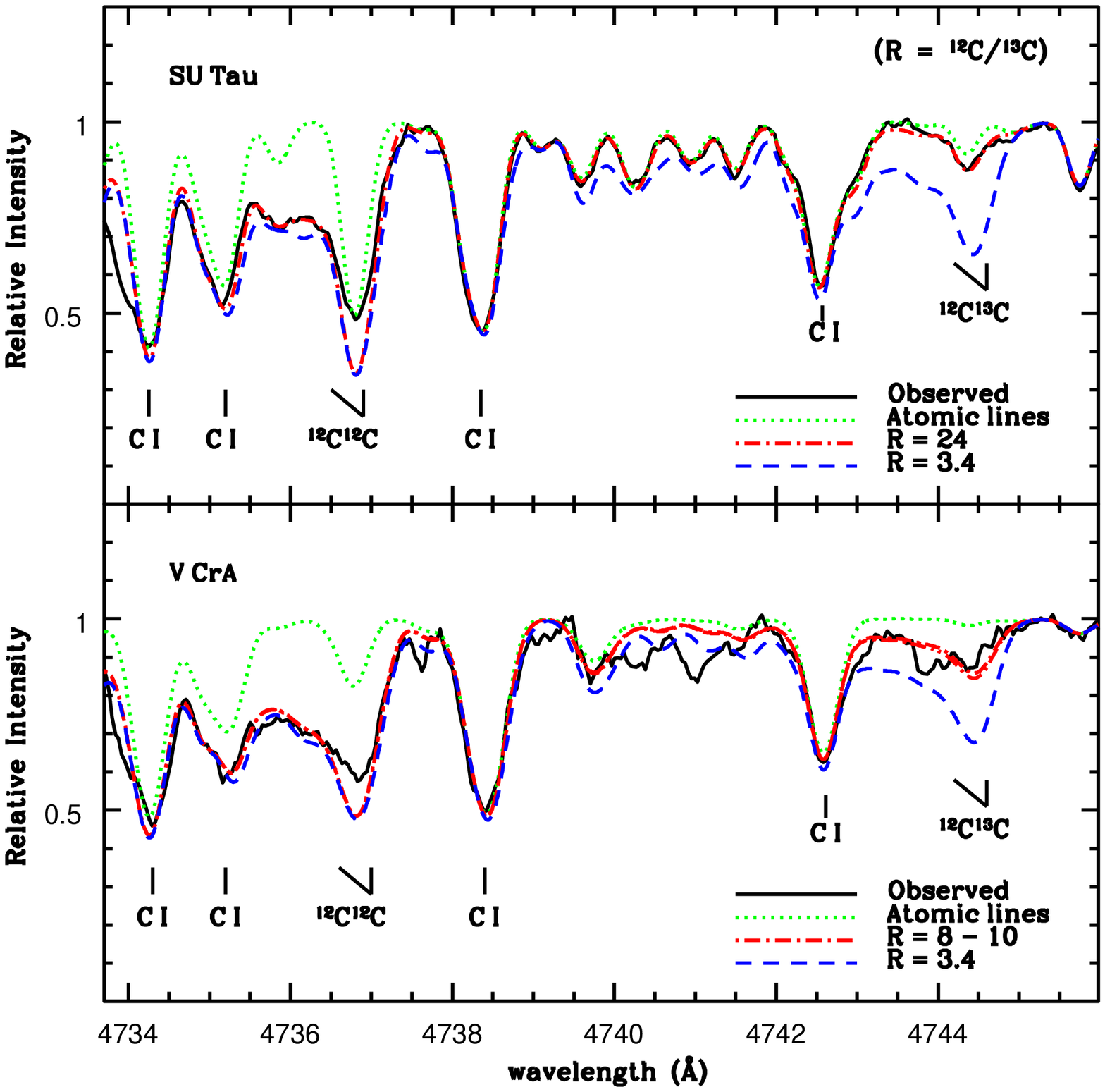}
\caption{Observed and synthetic spectra of the (1,0) C$_{2}$ bands for
SU\,Tau and V\,CrA. Synthetic spectra are plotted for the values
of the isotopic ratios (R) shown in the keys and for a spectrum
with just the atomic lines.  The positions of the key lines are also marked.
\label{}}
\end{figure}

\begin{figure}
\epsscale{1.0}
\plotone{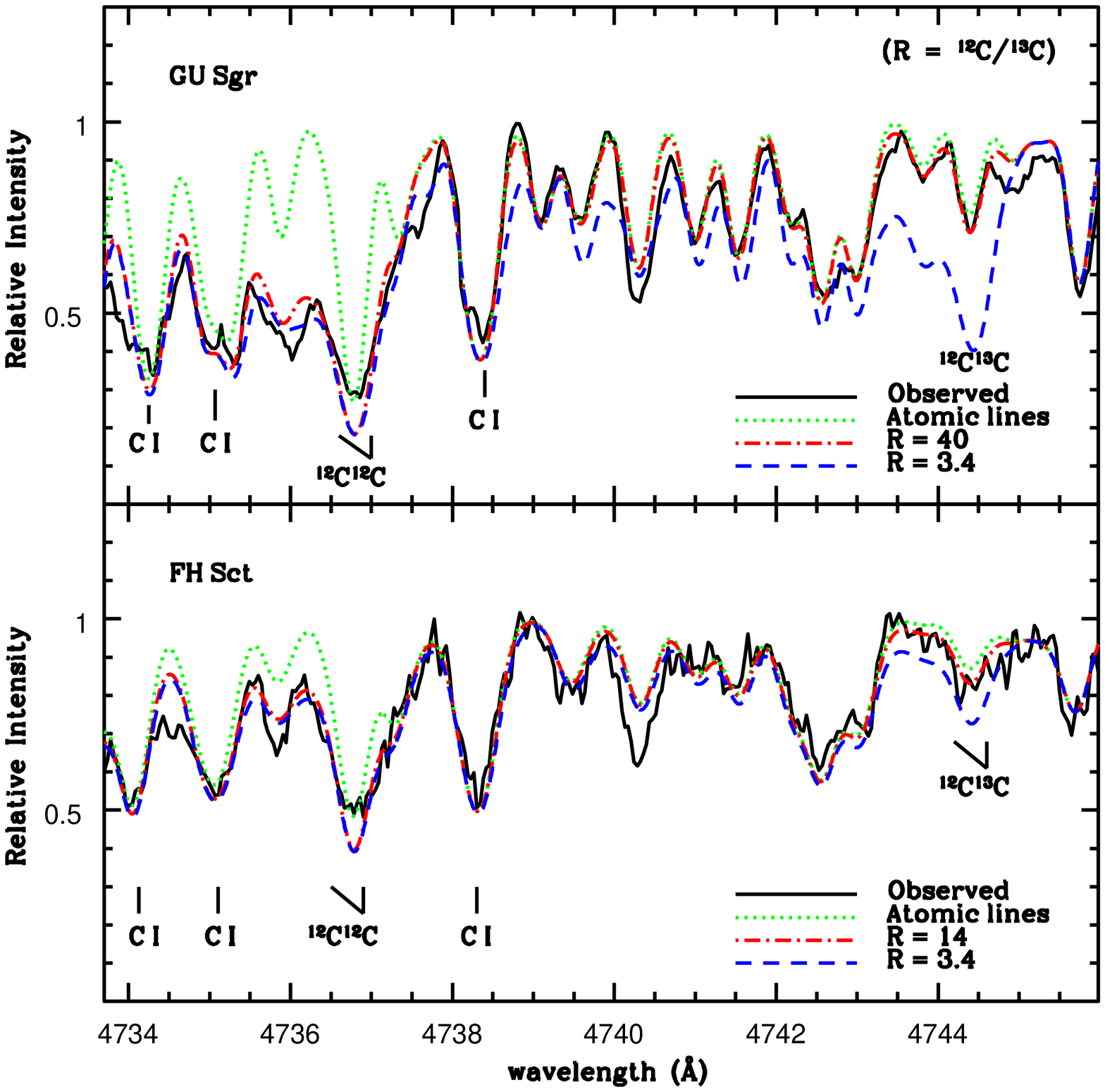}
\caption{Observed and synthetic spectra of the (1,0) C$_{2}$ bands for
GU\,Sgr and FH\,Sct. Synthetic spectra are plotted for the values
of the isotopic ratios (R) shown in the keys and for a spectrum
with just the atomic lines.  The positions of the key lines are also marked.
\label{}}
\end{figure}

\begin{figure}
\epsscale{1.0}
\plotone{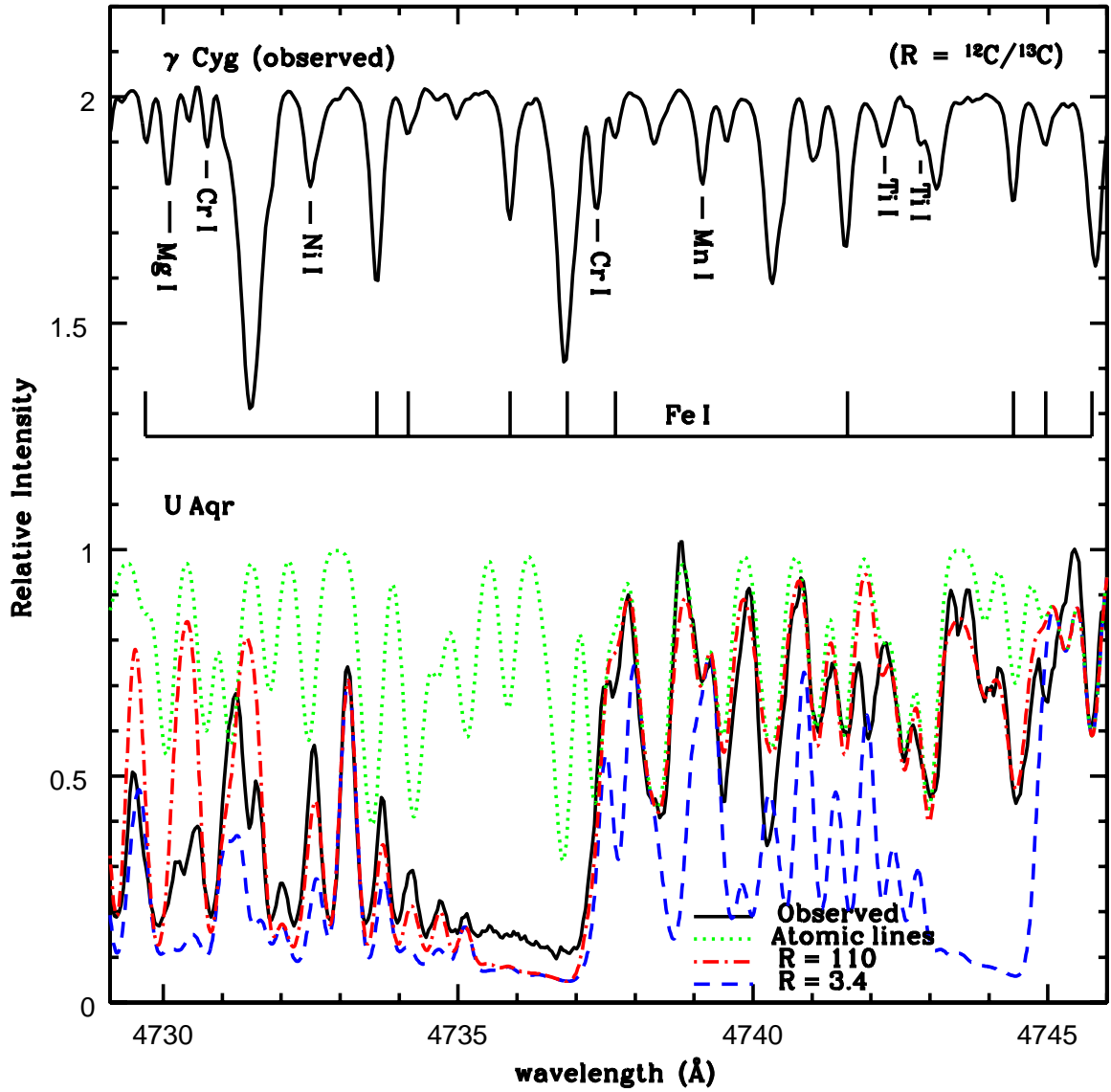}
\caption{Observed and synthetic spectra of the (1,0) C$_{2}$ bands for
U\,Aqr. Synthetic spectra are plotted for the values
of the isotopic ratios (R) shown in the keys and for a spectrum
with just the atomic lines.
The spectrum of $\gamma$\,Cyg is also plotted -- the positions of the
key lines are also marked.
\label{}}
\end{figure}

\begin{figure}
\epsscale{1.0}
\plotone{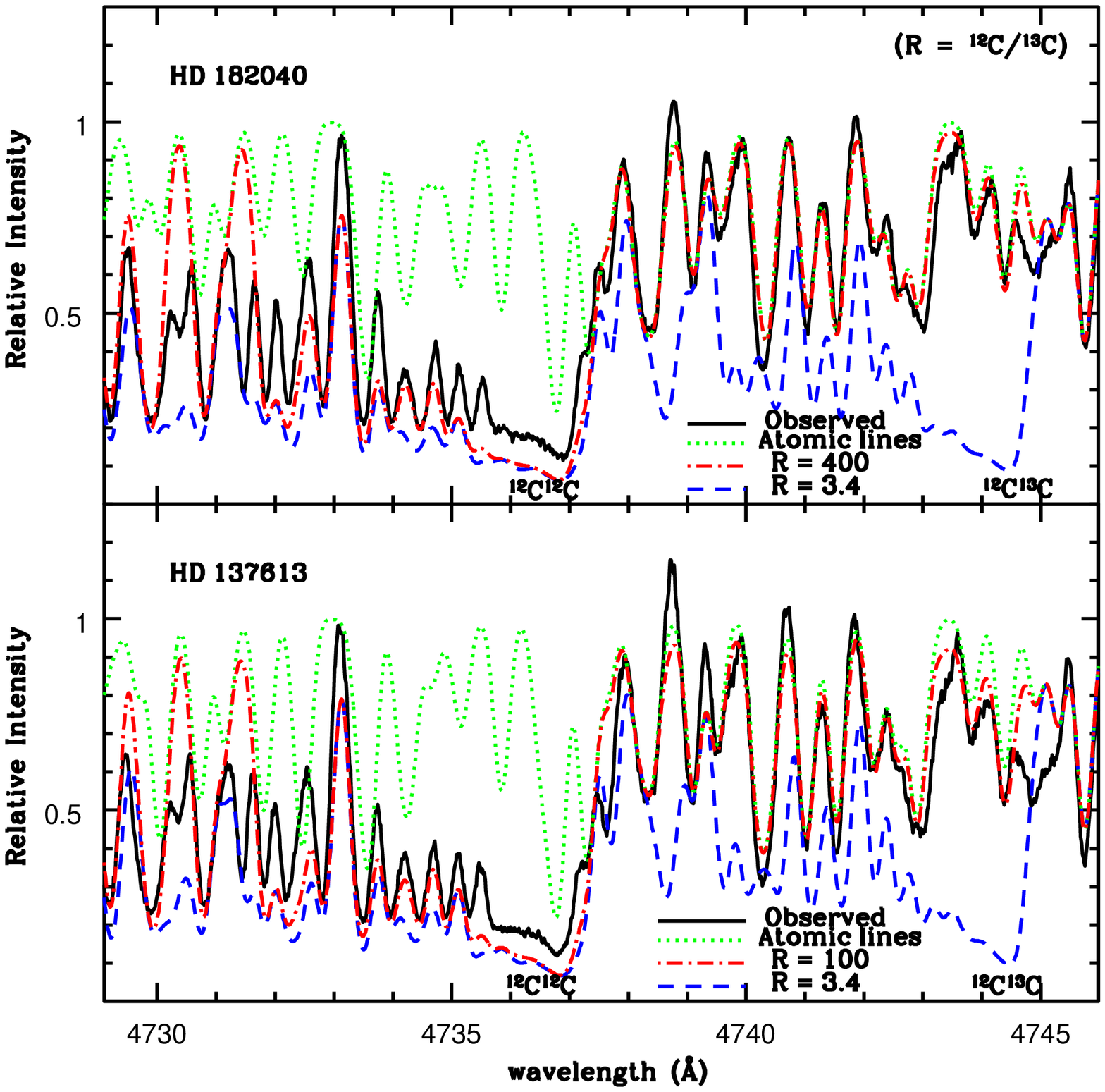}
\caption{Observed and synthetic spectra of the (1,0) C$_{2}$ bands for
HD\,182040 and HD\,137613. Synthetic spectra are plotted for the values
of the isotopic ratios (R) shown in the keys and for a spectrum
with just the atomic lines. The positions of the key lines are also marked.
\label{}}
\end{figure}

\begin{figure}
\epsscale{1.0}
\plotone{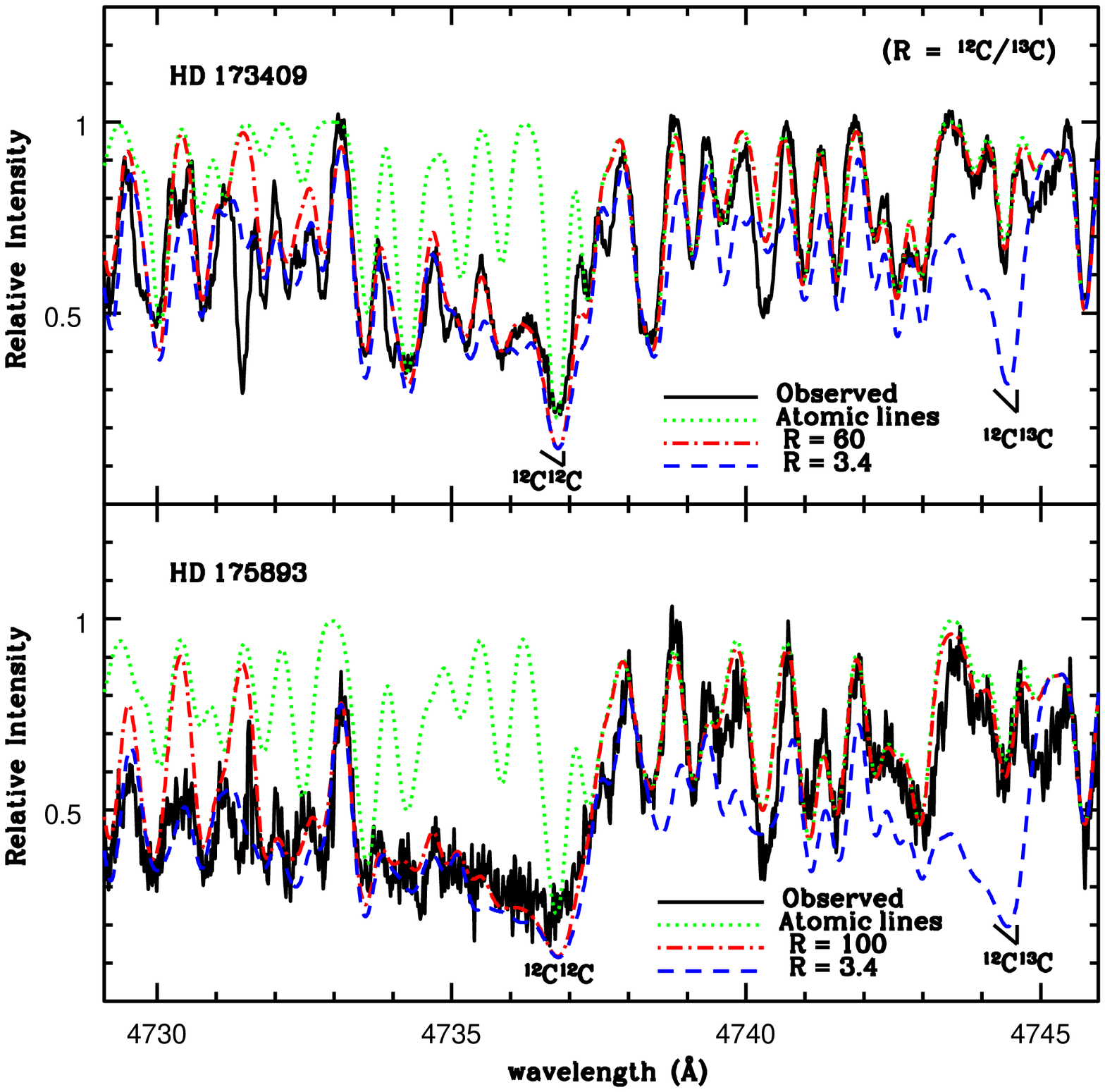}
\caption{Observed and synthetic spectra of the (1,0) C$_{2}$ bands for
HD\,173409 and HD\,175893. Synthetic spectra are plotted for the values
of the isotopic ratios (R) shown in the keys and for a spectrum
with just the atomic lines. The positions of the key lines are also marked.
\label{}}
\end{figure}

\clearpage
\begin{figure}
\epsscale{1.0}
\plotone{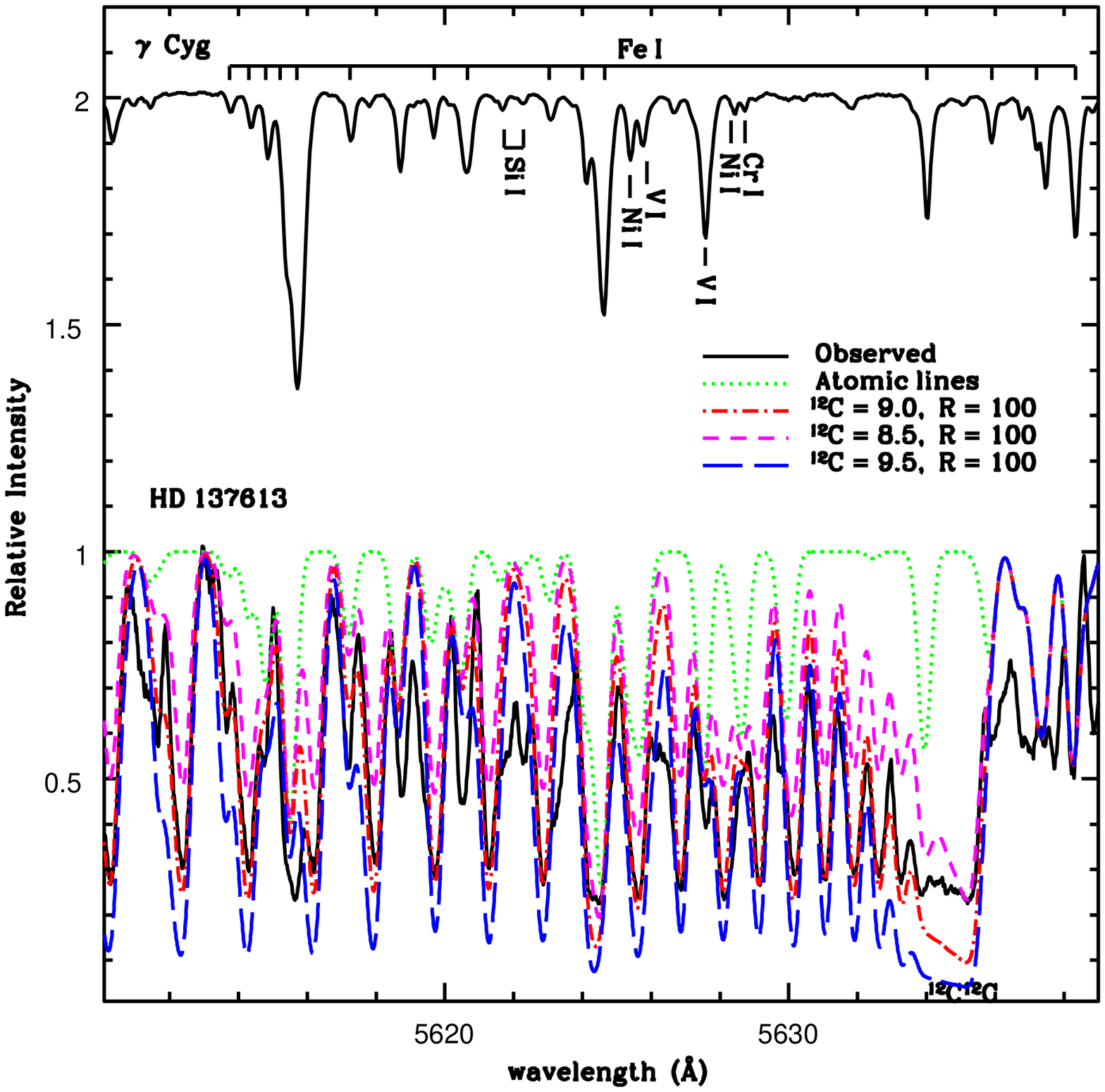}
\caption{Observed and synthetic spectra of the (0,\,1) C$_{2}$ band for
HD\,137613. Synthetic spectra are plotted for different
values of the C abundance -- see key on the figure.
The spectrum of the $\gamma$\,Cyg is
plotted with the positions of the key lines marked.
\label{}}
\end{figure}

\clearpage
\begin{figure}
\epsscale{1.0}
\plotone{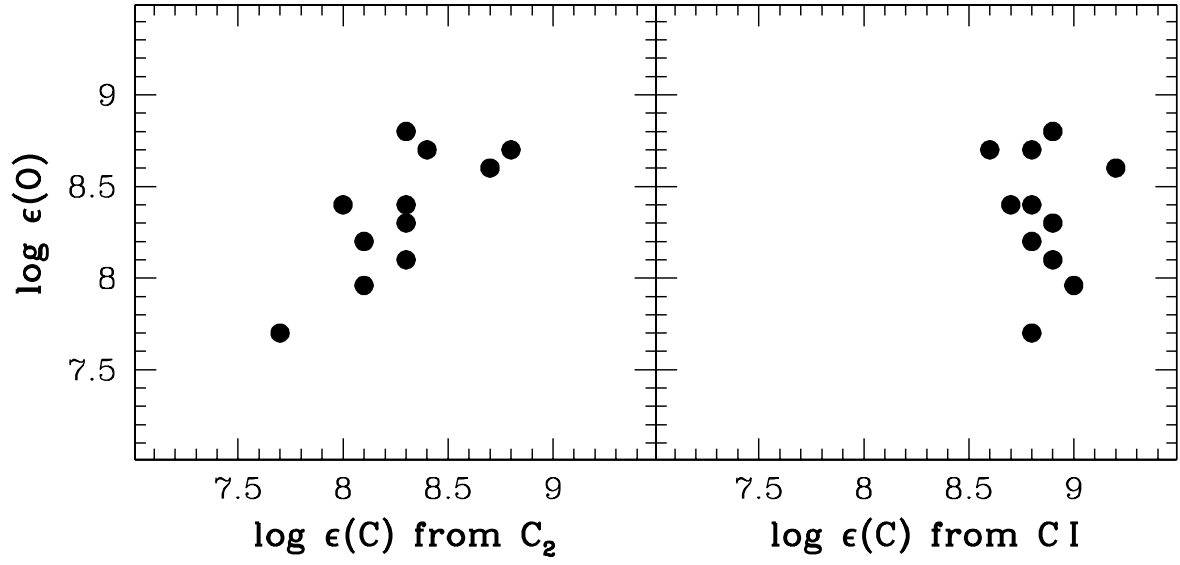}
\caption{The plot of log $\epsilon$(C), from C$_{2}$ bands and  C\,{\sc i}
lines versus log $\epsilon$(O) for RCB stars. The log $\epsilon$(O) and the
log $\epsilon$(C), from C\,{\sc i} lines, are from \citet{asplund00}.
\label{}}
\end{figure}

\clearpage
\begin{figure}
\epsscale{1.0}
\plotone{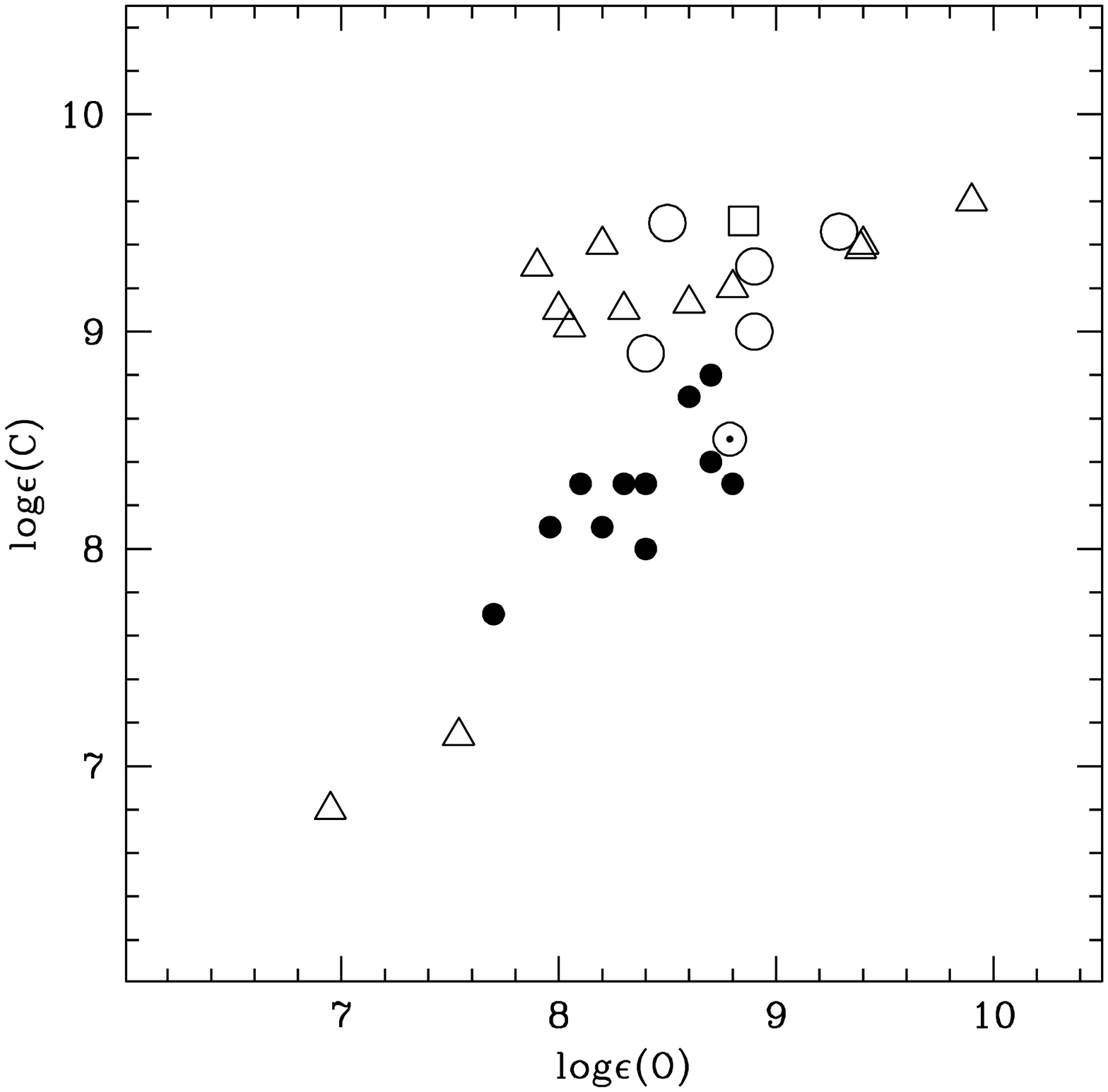}
\caption{The plot of log $\epsilon$(C) from C$_{2}$ bands
versus log $\epsilon$(O) \citep{asplund00}
for RCB stars. Also shown are the EHe stars.
Our sample of eleven RCBs are represented by filled circles.
Five cool EHes are represented by open circles
\citep{pandey01, 2006ApJ...638..454P,2006MNRAS.369.1677P}.
Twelve hot EHes are represented by open triangles
\citep{drilling98, jeffery97, jeffery98, jeffery99, pandey11}.
DY\,Cen, the hot minority RCB
\citep{jeffery93}
is represented by open square.
$\odot$ represents the Sun.
\label{}}
\end{figure}

\clearpage
\begin{table*}
\begin{center}
\caption{Log of the observations\,: the stars are listed in the decreasing order of their effective temperature from top to bottom.}
\begin{tabular}{cccccccccccc}
\tableline\tableline
Star &  Date of Observation &  V  & Obervatory & S/N Ratio \\
\tableline
V3795\,Sgr& 26 July 1996 & 11.2 & McDonald & 110\\
XX\,Cam & 17 November 2002 & 7.4 & McDonald & 200\\
VZ\,Sgr & 22 May 2007 & 10.2 & McDonald &  200 \\
UX\,Ant & 5 May 2007 & 12.8 &  McDonald & 120 \\
RS\,Tel & 28/29 May 2010 & 9.9 & VBT & 25 \\
R\,CrB  & 5 May 2007 & 6.0 & McDonald & 200 \\
V2552\,Oph & 22 May 2007 & 11.0 & McDonald & 128 \\
V854\,Cen & 24-27 May 2010/10 February 1999 & 7.25 & VBT/McDonald & 250 \\
V482\,Cyg & 23/24 May 2007 & 10.8 & McDonald & 152\\
SU\,Tau & 15 Nov 2002 & 9.8 & McDonald & 196 \\
V\,CrA & 6 September 2003 & 9.5  & McDonald & 137 \\
GU\,Sgr & 23 May 2007 & 11.1 & McDonald & 135\\
FH\,Sct & 24 May 2007 & 12.1 & McDonald & 87 \\
U\,Aqr & 23 July 1996 & 11.2 & McDonald & 125 \\
HD\,173409 & 27 May 2010 & 9.5 & VBT & 70 \\
HD\,182040 & 25 May 2010 & 7.0 & VBT & 110 \\
HD\,175893 & 25 May 2010 & 9.3 & VBT & 30 \\
HD\,137613 & 24 May 2010 & 7.5 & VBT & 90 \\
\tableline
\end{tabular}
\end{center}
\end{table*}

\clearpage
\begin{deluxetable}{cccccccc}
\tablecolumns{3}
\tablewidth{0pc}
\tablecaption{Molecular line list for the (1,\,0) C$_2$ swan band - $^{12}$C$^{12}$C lines}
\tablehead{
\colhead{Wavelength (\AA)} & \colhead{$J^{''}$} & \colhead{$\chi$ (eV)}   & \colhead{log $gf$ }}
\startdata
   4692.348&      28.0&        0.342&      -0.270\\
   4692.485&      28.0&        0.342&      -0.270\\
   4692.548&      27.0&        0.342&      -0.286\\
   4692.679&      26.0&        0.342&      -0.302\\
   4692.794&      61.0&        0.940&       0.064\\
   4692.838&      61.0&        0.940&       0.064\\
   4692.848&      62.0&        0.940&       0.071\\
   4692.931&      60.0&        0.940&       0.057\\
\enddata
\end{deluxetable}

\clearpage
\begin{deluxetable}{cccccccc}
\tablecolumns{3}
\tablewidth{0pc}
\tablecaption{Molecular line list for the (1,\,0) C$_2$ swan band - $^{12}$C$^{13}$C lines}
\tablehead{
\colhead{Wavelength (\AA)} & \colhead{$J^{"}$} & \colhead{$\chi$ (eV)}   & \colhead{log $gf$ }}
\startdata
   4700.760&        28.0&        0.342&      -0.270\\
   4701.749&        28.0&        0.342&      -0.270\\
   4701.736&        27.0&        0.342&      -0.286\\
   4701.789&        26.0&        0.342&      -0.302\\
   4703.775&        61.0&        0.940&       0.064\\
   4702.031&        61.0&        0.940&       0.064\\
   4702.111&        62.0&        0.940&       0.071\\
   4702.041&        60.0&        0.940&       0.057\\
\enddata
\end{deluxetable}

\clearpage
\begin{deluxetable}{cccccccc}
\tablecolumns{3}
\tablewidth{0pc}
\tablecaption{Molecular line list for the (0,\,0) C$_2$ swan band - $^{12}$C$^{12}$C lines}
\tablehead{
\colhead{Wavelength (\AA)} & \colhead{$J^{"}$} & \colhead{$\chi$ (eV)}   & \colhead{log $gf$ }}
\startdata
   5129.601&        47.0&        0.640&       0.465\\
   5129.601&        48.0&        0.640&       0.474\\
   5129.601&        47.0&        0.640&       0.465\\
   5129.601&        48.0&        0.640&       0.474\\
   5129.694&        46.0&        0.640&       0.455\\
   5129.694&        46.0&        0.640&       0.455\\
   5130.272&        19.0&        0.259&       0.078\\
   5130.418&        18.0&        0.259&       0.055\\
\enddata
\end{deluxetable}

\clearpage
\begin{deluxetable}{cccccccc}
\tablecolumns{3}
\tablewidth{0pc}
\tablecaption{Molecular line list for the (0,\,0) C$_2$ swan band - $^{12}$C$^{13}$C lines}
\tablehead{
\colhead{Wavelength (\AA)} & \colhead{$J^{"}$} & \colhead{$\chi$ (eV)}   & \colhead{log $gf$ }}
\startdata
   5132.755&        47.0&        0.640&       0.465\\
   5132.877&        48.0&        0.640&       0.474\\
   5131.091&        47.0&        0.640&       0.465\\
   5131.171&        48.0&        0.640&       0.474\\
   5132.735&        46.0&        0.640&       0.455\\
   5131.101&        46.0&        0.640&       0.455\\
   5131.816&        19.0&        0.259&       0.078\\
   5131.879&        18.0&        0.259&       0.055\\
\enddata
\end{deluxetable}

\clearpage
\begin{deluxetable}{cccccccc}
\tablecolumns{3}
\tablewidth{0pc}
\tablecaption{Molecular line list for the (0,\,1) C$_2$ swan band - $^{12}$C$^{12}$C lines}
\tablehead{
\colhead{Wavelength (\AA)} & \colhead{$J^{"}$} & \colhead{$\chi$ (eV)}   & \colhead{log $gf$ }}
\startdata
   5587.533&        21.0&        0.474&      -0.360\\
   5587.684&        20.0&        0.474&      -0.381\\
   5587.876&        19.0&        0.475&      -0.403\\
   5588.125&        45.0&        0.800&      -0.035\\
   5588.125&        45.0&        0.800&      -0.035\\
   5588.125&        46.0&        0.800&      -0.025\\
   5588.125&        46.0&        0.800&      -0.025\\
   5588.263&        44.0&        0.800&      -0.044\\
\enddata
\end{deluxetable}

\clearpage
\begin{deluxetable}{cccccccc}
\tablecolumns{3}
\tablewidth{0pc}
\tablecaption{Molecular line list for the (0,\,1) C$_2$ swan band - $^{12}$C$^{13}$C lines}
\tablehead{
\colhead{Wavelength (\AA)} & \colhead{$J^{"}$} & \colhead{$\chi$ (eV)}   & \colhead{log $gf$ }}
\startdata
   5579.970&        21.0&        0.474&      -0.360\\
   5580.008&        20.0&        0.474&      -0.381\\
   5580.092&        19.0&        0.475&      -0.403\\
   5582.309&        45.0&        0.800&      -0.035\\
   5580.424&        45.0&        0.800&      -0.035\\
   5582.465&        46.0&        0.800&      -0.025\\
   5580.532&        46.0&        0.800&      -0.025\\
   5582.300&        44.0&        0.800&      -0.044\\
\enddata
\end{deluxetable}

\clearpage
\hoffset -0.4 in
\begin{deluxetable}{ccccccccccccccccc}
\tablecolumns{7}
\tablewidth{0pc}
\tablecaption{The atomic line list used in the syntheses of the (1,\,0) C$_{2}$
swan band region with the individual estimates of the log $gf$-values from the
$\gamma$\,Cyg, Sun, and Arcturus spectra and the adopted log $gf$-values.}
\tablehead{
\colhead{Line}  & \colhead{$\chi$ (eV)} & \colhead{log $gf_{\rm \gamma\,Cyg}$\tablenotemark{a}} & \colhead{log $gf_{\rm Sun}$\tablenotemark{b}} & \colhead{log $gf_{\rm Arcturus}$\tablenotemark{c}} & \colhead{log $gf_{\rm Source}$} & \colhead{Source} & \colhead{log $gf_{\rm adopted}$}}
\startdata
Fe\,{\sc i}\,$\lambda$4729.018&4.07&-1.72&-1.60&-1.70&-1.61&NIST&-1.72\\
Ni\,{\sc i}\,$\lambda$4729.280&4.10&-2.00&-1.20&\nodata&-1.20&NIST&-2.00\\
Fe\,{\sc i}\,$\lambda$4729.676&3.40&-2.36&-2.32&-2.50&-2.42&NIST&-2.36\\
Mg\,{\sc i}\,$\lambda$4730.028&4.34&-2.49&-2.42&-2.49&-2.34&NIST&-2.49\\
Cr\,{\sc i}\,$\lambda$4730.710&3.08&-0.48&-0.38&-0.38&-0.19&NIST&-0.48\\
Ti\,{\sc i}\,$\lambda$4731.165&2.17&\nodata&-0.51&-0.51&-0.41&Kurucz&-0.51\\
Ni\,{\sc i}\,$\lambda$4731.798&3.83&\nodata&-0.93&-1.10&-0.85&NIST&-0.93\\
Ni\,{\sc i}\,$\lambda$4732.457&4.10&-0.59&-0.55&\nodata&-0.55&NIST&-0.59\\
\enddata
\tablenotetext{a}{MARCS Model atmosphere with atmospheric parameters and abudances from \citet{luck81}.}
\tablenotetext{b}{MARCS Model atmosphere with atmospheric parameters and abudances from \citet{asplund09}.}
\tablenotetext{c}{MARCS Model atmosphere with atmospheric parameters and abudances from \citet{peterson93}.}
\end{deluxetable}

\begin{deluxetable}{cccccccccccc}
\tablecaption{The Fe abundances for RCB and HdC stars.}
\tablewidth{0pt}
\tablehead{
\colhead{Star} & \colhead{log$\epsilon$(Fe)\tablenotemark{a}} &
\colhead{log$\epsilon$(Fe)\tablenotemark{b}} &
\colhead{log$\epsilon$(Fe)\tablenotemark{c}}}
\startdata
V3795\,Sgr&   5.7(0.3)(3) &    5.6 &      $<$6.0\\
XX\,Cam &     6.8(0.3)(5) &    6.8 &      6.8 \\
VZ\,Sgr &     6.1(0.3)(4) &    5.8 &      7.2 \\
UX\,Ant &     6.2(0.15)(5)&    6.2   &    7.0 \\
RS\,Tel &     6.5(0.2)(3) &    6.4  &     6.9\\
R\,CrB  &     6.6(0.2)(5)&     6.5&       7.0    \\
V2552\,Oph &  6.6(0.2)(4) &    6.4  &     6.9    \\
V854\,Cen &   5.0(0.3)(3)&     5.0&       6.5    \\
V482\,Cyg &   6.7(0.15)(5) &   6.7 &      6.9    \\
SU\,Tau  &    6.1(0.3)(4)&     6.1 &      6.5    \\
V\,CrA   &    5.5(0.1)(3) &    5.5  &     6.6    \\
GU\,Sgr  &    6.5(0.25)(6) &   6.3 &      6.8    \\
FH\,Sct  &    6.4(0.15)(5) &   6.3   &    6.8    \\
U\,Aqr   &    6.5(0.30)(8) &   \nodata &  7.3\\
HD\,173409&   6.6(0.3)(10) &   6.8&       6.6\\
HD\,182040&   6.6(0.25)(9) &   6.9 &      6.4 \\
HD\,175893&   6.7(0.2)(5)  &   6.8&       6.7\\
HD\,137613&   6.8(0.20)(9) &   6.6 &      6.7\\
\enddata
\tablenotetext{a}{from 4700\,\AA\ region, the values in parentheses are
the standard deviation and the number of lines used, respectively.}
\tablenotetext{b}{from \citet{asplund00} for RCB stars and from \citet{warner67} for HdC stars.}
\tablenotetext{c}{from 4744.4\,\AA\ line, assuming it to be entirely Fe\,{\sc i}.}
\end{deluxetable}

\clearpage
\thispagestyle{empty}
\begin{landscape}
\voffset 1.5 in
\begin{deluxetable}{ccccccccccccccc}
\tabletypesize{\scriptsize}
\tablecolumns{14}
\tablewidth{-40pc}
\tablecaption{The derived carbon abundances for RCB stars from (0,\,0), (0,\,1) and (1,\,0) C$_{2}$ bands.}
\tablehead{
\colhead{}    &  \multicolumn{3}{c}{log$\epsilon$(C) from (0,\,1) C$_{2}$ band} & \colhead{} & \multicolumn{3}{c}{log$\epsilon$(C) from (0,\,0)} & \colhead{} & \multicolumn{3}{c}{log$\epsilon$(C) from (1,\,0)} & \colhead{log$\epsilon$(C) from C\,{\sc i} lines}\\
\cline{2-4} \cline{6-8} \cline{10-12} \\
\colhead{stars} & \colhead{C/He = 0.3\%}   & \colhead{C/He = 1.0\%}    & \colhead{C/He = 3.0\%} & \colhead{} & \colhead{C/He = 0.3\%}   & \colhead{C/He = 1.0\%}    & \colhead{C/He = 3.0\%} & \colhead{} &      \colhead{C/He = 0.3\%}   & \colhead{C/He = 1.0\%}    & \colhead{C/He = 3.0\%} & \colhead{C/He = 1.0\%} \\
&log $\epsilon$(C) = 9.0 & log $\epsilon$(C) = 9.5 & log $\epsilon$(C) = 10.0 & & log $\epsilon$(C) = 9.0 & log $\epsilon$(C) = 9.5 & log $\epsilon$(C) = 10.0 &
& log $\epsilon$(C) = 9.0 & log $\epsilon$(C) = 9.5 & log $\epsilon$(C) = 10.0 & log $\epsilon$(C) = 9.5}
\startdata
VZ\,Sgr & 9.0  & 8.9 &
 8.8 & & 9.0 & 8.8 & 8.7 &
& 9.0 & 8.8 &  8.6 & 8.9\\
UX\,Ant & 8.4 & 8.3 &
 8.2 & & 8.2 & 8.1 & 8.0 &
& 8.4 & 8.1  & 8.0 & 8.7 \\
RS\,Tel & \nodata & \nodata &
\nodata & & 8.4 & 8.3 & 8.3 &
&  8.7 & 8.5 & 8.4 & 8.7\\
R\,CrB &9.0 & 8.8 &
 8.8 & & 8.9 & 8.6 & 8.6 &
& 9.0 & 8.8 & 8.7 & 8.9\\
V2552\,Oph&8.3& 8.1 &
 8.2 & & 8.1 & 8.1 &  8.1 &
& 8.1 & 8.0 & 7.9 & 8.7\\
V854\,Cen&8.4& 8.3 &
 8.3 & &8.4  & 8.3 &  8.2 &
& \nodata & 8.5 & 8.2 & 8.8\\
V482\,Cyg&8.4 & 8.3 &
 8.3 & & 8.2 & 8.1 & 8.1 &
& 8.2 & 8.2 & 8.1 & 8.9\\
SU\,Tau&8.1 & 8.0 &
8.0 & & 7.8 & 7.8 & 7.8 &
& 7.8 & 7.7 & 7.7 & 8.6\\
V\,CrA&8.5& 8.4 &
8.3 & & 8.3 & 8.2 & 8.2 &
& 8.5 & 8.4 & 8.3 & 8.6\\
GU\,Sgr&8.2& 8.1 &
8.1 & & 8.1 & 8.1 & 8.1 &
& 8.2 & 8.1 & 8.0 & 8.9\\
FH\,Sct&7.8 & 7.7 &
7.7 & & 7.8 & 7.8 & 7.7 &
& 7.7 & 7.7 & 7.6 & 8.9\\
\enddata
\end{deluxetable}
\end{landscape}

\clearpage
\thispagestyle{empty}
\begin{landscape}
\voffset 1.4 in
\begin{deluxetable}{ccccccccccccccc}
\tabletypesize{\normalsize}
\tablecolumns{14}
\tablewidth{-40pc}
\tablecaption{The derived carbon abundances for HdC stars and RCB star U\,Aqr from (0,\,1) and (1,\,0) C$_{2}$ bands.}
\tablehead{
\colhead{}    &  \multicolumn{3}{c}{log$\epsilon$(C) from (0,\,1) C$_{2}$ band} & \colhead{} &  \multicolumn{3}{c}{log$\epsilon$(C) from (1,\,0)} & \colhead{log$\epsilon$(C) from C\,{\sc i} lines}\\
\cline{2-4} \cline{6-8} \\
\colhead{stars} & \colhead{C/He = 0.1\%}   & \colhead{C/He = 1.0\%}    & \colhead{C/He = 10\%} & \colhead{}  &  \colhead{C/He = 0.1\%}   & \colhead{C/He = 1.0\%}    & \colhead{C/He = 10\%} & \colhead{C/He = 1.0\%}\\
& log $\epsilon$(C) = 8.5 & log $\epsilon$(C) = 9.5  & log $\epsilon$(C) = 10.5 & & log $\epsilon$(C) = 8.5 & log $\epsilon$(C) = 9.5 & log $\epsilon$(C) = 10.5 & log $\epsilon$(C) = 9.5 }
\startdata
HD\,173409 & \nodata & 8.7 &
\nodata & & \nodata & 8.7 & \nodata & 8.6\\
HD\,182040 & 8.8 & 9.0 &
8.9 & & 8.8 & 9.0 & 8.9 & 9.0\\
HD\,175893 & 8.9 & 9.0 &
8.9 & & 8.8 & 8.9 & 8.8 & 8.5 \\
HD\,137613 & 8.8 & 9.0 &
8.9 & & 8.8 & 9.0 & 8.9 & 8.5\\
U\,Aqr\tablenotemark{*}&\nodata&9.2&\nodata&&
\nodata&9.2&\nodata & 8.9\\
\enddata
\tablenotetext{*}{Adopted ($T_{\rm eff}$, log \textit{g}) = (5400, 0.5).
If ($T_{\rm eff}$, log \textit{g}) = (6000, 0.5) is adopted,
the derived carbon abundance is about 10.4.}
\end{deluxetable}
\end{landscape}

\clearpage
\begin{deluxetable}{cccccccccccc}
\tablecaption{The adopted stellar parameters and the $^{12}$C/$^{13}$C ratios for the analysed stars.}
\tablewidth{0pt}
\tablehead{
\colhead{Star} & \colhead{($T_{\rm eff}$[K], log $g$[cgs], $\xi_{t}$[km\,s$^{-1}$])} & \colhead{$^{12}$C/$^{13}$C Ratio\tablenotemark{a}} & \colhead{$^{12}$C/$^{13}$C Ratio}
}
\startdata
VZ\,Sgr & (7000, 0.5, 7.0)&  3\,-\,6&\nodata\\
UX\,Ant & (6750, 0.5, 5.0)& 14\,-\,20  &\nodata\\
RS\,Tel & (6750, 1.0, 8.0) & $>$60  & \nodata\\
R\,CrB  & (6750, 0.5, 7.0) & $>$40  & $>$40\tablenotemark{b} \\
V2552\,Oph & (6750, 0.5, 7.0) & $>$8 & \nodata\\
V854\,Cen &(6750, 0.0, 6.0)& 16\,-\,24 & \nodata \\
V482\,Cyg & (6500, 0.5, 4.0)& $>$100 & \nodata\\
SU\,Tau & (6500, 0.5, 7.0)&$>$24 & \nodata\\
V\,CrA & (6500, 0.5, 7.0) & 8\,-\,10 & 4\,-\,10\tablenotemark{c}\\
GU\,Sgr & (6250, 0.5, 7.0) & $>$40 & \nodata\\
FH\,Sct & (6250, 0.0, 6.0) & $>$14 & \nodata\\
U\,Aqr & (5400, 0.5, 5.0) & 110\,-\,120 & \nodata \\
HD\,173409 & (6100, 0.5, 6.0) &  $>$60 & \nodata\\
HD\,182040 & (5400, 0.5, 6.5) &  $>$400 & $>$100\tablenotemark{d,e} \\
HD\,175893 & (5400, 0.5, 6.0) & $>$100 & \nodata \\
HD\,137613 & (5400, 0.5, 6.5) & $>$100  & $>$500\tablenotemark{e} \enddata
\tablecomments{$^{a}$ Present work, $^{b}$ Cottrell \& Lambert\,(1982), $^{c}$ Rao \& Lambert\,(2008), \\
$^{d}$ Climenhaga (1960), $^{e}$ Fujita \& Tsuji\,(1997).}
\end{deluxetable}

\clearpage
\thispagestyle{empty}
\begin{landscape}
\hoffset -0.5 in
\voffset 1.4 in
\begin{deluxetable}{cccccccccccc}
\tablecaption{The log $\epsilon$(C) from (0,0) and (0,1) C$_{2}$ bands, except for RS\,Tel which is only from
(0,\,0) C$_{2}$ band, with C/He ratio 1\%. }
\tablewidth{0pt}
\tablehead{
\colhead{Star} & \colhead{($T_{\rm eff}$, log $g$)} &  \colhead{($T_{\rm eff}$-250, log $g$)} & \colhead{($T_{\rm eff}$, log $g$)} &
\colhead{($T_{\rm eff}$+250, log $g$)} & \colhead{($T_{\rm eff}$, log $g$-0.5)}
& \colhead{($T_{\rm eff}$, log $g$)} & ($T_{\rm eff}$, log $g$+0.5) & log $\epsilon$ (C\,{\sc i})}
\startdata
VZ\,Sgr& (7000, 0.5) &  8.3&8.8&\nodata&\nodata&8.8& 8.4 & 8.9\\
UX\,Ant& (6750, 0.5) & 7.9&8.3&\nodata&\nodata&8.3&8.0 & 8.7\\
RS\,Tel& (6750, 1.0) & 8.1& 8.3&\nodata& 8.6& 8.3& 8.2& 8.7\\
R\,CrB&  (6750, 0.5) &8.3&8.8&\nodata&\nodata&8.8&8.4 & 8.9\\
V2552\,Oph&(6750, 0.5) & 7.9&8.1&\nodata& 8.5&8.1&8.0 & 8.7\\
V854\,Cen& (6750, 0.0) &7.8&8.3&\nodata&\nodata&8.3&7.9 & 8.8\\
V482\,Cyg&(6500, 0.5)&  8.1&8.3&8.7&8.6&8.3&8.2 & 8.9\\
SU\,Tau&(6500, 0.5)& 7.8&8.0&8.4&8.3&8.0&7.9 & 8.6\\
V\,CrA&(6500, 0.5) &8.2&8.4&8.8&8.7&8.4&8.3 & 8.6\\
GU\,Sgr& (6250, 0.5) & 7.9&8.1&8.4&8.3&8.1&8.1 & 8.9\\
FH\,Sct& (6250, 0.0)&7.5&7.7&8.0&7.9&7.7&7.6 & 8.9\\
\tablebreak
\thispagestyle{empty}
U\,Aqr& (5400, 0.5) & 8.9 & 9.2 & 9.5& \nodata&9.2&\nodata&8.9\\
HD\,173409 & (6100, 0.5) & 8.4 &  8.7 &  9.0& \nodata&8.7&\nodata&8.6\\
HD\,182040 & (5400, 0.5) & 8.7 &  9.0 &  9.3 &\nodata&9.0&\nodata&9.0\\
HD\,175893 & (5400, 0.5) & 8.7 &  9.0 &  9.3 &\nodata&9.0&\nodata&8.5\\
HD\,137613 & (5400, 0.5) & 8.7 &  9.0 &  9.3 &\nodata&9.0&\nodata&8.5\\
\enddata
\end{deluxetable}
\end{landscape}

\end{document}